\documentclass{article}

\usepackage{arxiv}

\usepackage[utf8]{inputenc} 
\usepackage[T1]{fontenc}    
\usepackage{hyperref}       
\usepackage{url}            
\usepackage{booktabs}       
\usepackage{amsfonts, amsmath, amssymb}       
\usepackage{nicefrac}       
\usepackage{microtype}      
\usepackage{lipsum}		
\usepackage{graphicx}
\usepackage{natbib}
\usepackage{doi}
\usepackage{fancyhdr, caption, subcaption, multirow}
\usepackage{indentfirst}
\usepackage{setspace}

\newcommand{\Adj}{\textrm{Adj}}

\newcommand{\expect}[1]     {\mathbb{E}\left[ #1 \right]}
\newcommand{\stdev}[1]     {\sigma\left[ #1 \right]}
\newcommand{\Cov}[2]{\mathrm{Cov}\!\left[#1,#2\right]}
\DeclareMathAlphabet{\mathbbold}{U}{bbold}{m}{n}

\title{Rapid neural network prediction of linear block copolymer free energies}

\author{Ian Chen \\
	Department of Materials Science and Engineering\\
	Massachusetts Institute of Technology\\
	Cambridge, MA 02139 \\
	\texttt{ischen@mit.edu} \\
	\And
	Alfredo Alexander-Katz \\
	Department of Materials Science and Engineering\\
    Massachusetts Institute of Technology\\
	Cambridge, MA 02139 \\
	\texttt{aalexand@mit.edu} \\
}

\renewcommand{\undertitle}{A Preprint accompanying APS March Meeting 2026 presentation}
\renewcommand{\shorttitle}{Rapid NN prediction of BCP free energies}

\hypersetup{
pdftitle={Rapid neural network prediction of linear copolymer melt free energies},
pdfsubject={},
pdfauthor={Ian Chen, Alfredo Alexander-Katz},
pdfkeywords={block copolymers, machine learning, free energy},
}

\begin{document}
\maketitle

\begin{abstract}
Free energies are fundamental quantities governing phase behavior and thermodynamic stability in polymer systems, yet their accurate computation often requires extensive simulations and post-processing techniques such as the Bennett Acceptance Ratio (BAR). While BAR provides reliable estimates when applied between closely related thermodynamic states, evaluating free energies across large changes in interaction strength typically requires a sequence of intermediate simulations to maintain sufficient phase-space overlap, substantially increasing computational cost. In this work we develop a machine learning framework for rapidly predicting excess free energies of linear diblock copolymer systems from simulation-derived energetic descriptors. Using dissipative particle dynamics simulations of freely-jointed chain polymers, we construct a dataset of per-chain energetic statistics—including heterogeneous interaction energies, homogeneous interaction energies, and bonded spring energies—and train feed-forward neural networks to learn the relationship between these descriptors and free energies computed using a stratified BAR procedure. The resulting models accurately reproduce the reference free energies across a range of chain lengths, compositions, and densities, including polymer architectures held out from training. In regimes where direct, brute-force BAR estimates become unreliable due to poor phase-space overlap, the neural network predictions remain consistent with the reference values. These results demonstrate that physically informed machine learning models can serve as efficient surrogates for expensive free-energy calculations and provide a promising approach for accelerating thermodynamic analysis of polymer systems.
\end{abstract}


\section{Background and motivation}

Block copolymers (BCPs), polymers consisting of ``blocks'' of two or more distinct monomer chemistries, are simple molecules yet are able to self-assemble into highly ordered microstructures. \cite{flory1941thermodynamics, matsen1996origins, klok2001supramolecular} These microstructures have paramount importance and applications in a wide variety of domains. For instance, BCPs can self-assemble into water-soluble micelles for drug delivery, allowing hydrophobic drugs and/or small molecules to be transferred into hydrophilic media efficiently. \cite{kataoka2012block, kwon2006amphiphilic, gaucher2005block} Specialized microstructures can also form materials with unique physical properties; lamellae, a common BCP microstructure, are mechanically anisotropic. Starting with the insights provided by Flory in the mid-20th century, understanding the behavior of such polymer chains has gradually increased, with extensive and deliberate work invested in determining phase diagrams and macroscopic properties of these materials in both experimental and computational settings. \cite{flory1941thermodynamics, matsen1996origins, detcheverry2008monte} For diblock copolymers (dBCPs), or BCPs with exactly two chemistries, the phase diagram was first computed in the 1980s, with four primary ordered phases: lamellae, gyroids, cylinders, and spheres. \cite{flory1953principles} 

Unsurprisingly, several parameters increase the complexity of BCP phase diagrams. For dBCPs, the chemistries of the two blocks, as well as the phase fraction of each chemistry (i.e., how long one block is relative to the other), are primary parameters governing self-assembly. \cite{bates1999block, li2010solvent} Increasing the number of blocks, chemistries, and possible phase fractions rapidly expands the parameter space, along with an equally large output space for the resulting phases. Furthermore, a large phase output space corresponds to microstructures with a wide spectrum of potential properties. Measuring the properties of a wide range of BCP formulations, while combinatorially unfeasible in experimental settings, has been attempted computationally, often with the assistance of machine learning. \cite{tu2020machine, aoyagi2021deep, fang2025universal} However, much of the current research focuses on predicting fundamental polymer properties such as glass-transition temperature and physical moduli, largely due to the availability of experimental datasets for real-world validation. \cite{kim2021polymer, mysona2024machine, chang2025mechanical}


Machine learning (ML) is a broad term encompassing a wide range of methods, from linear regression to large-language models (LLMs), and generally describes programs that learn from data to generate desired outputs. As discussed previously, the difficulty of predicting BCP behavior as the complexity of the parameter space increases has motivated the use of ML methods to address these challenges. More specifically, training neural networks (NNs) to predict polymer behavior and properties has become a primary application of ML in polymer science. While some polymer properties exhibit approximately linear relations (e.g., free energy changes in the low $\chi$ regime), the majority of relevant properties are nonlinear, necessitating neural networks capable of learning nonlinear relationships. \cite{mysona2024machine, chang2025mechanical} Existing NN applications to BCPs primarily fall into two categories: (1) predicting macroscopic properties (mechanical, optical, etc.) of real-world polymers, and (2) predicting thermodynamic and phase behavior of theoretical polymer chain models.

We first consider NN applications to real-world polymers, as this area is more established and includes a wide variety of NN architectures used to predict polymer properties. Neural networks have been widely applied to model the highly nonlinear structure–property relationships that characterize polymer systems, enabling prediction of experimentally measurable macroscopic properties such as glass transition temperature ($T_g$), melting temperature ($T_m$), thermal conductivity, dielectric constant, elastic modulus, and solubility parameters. \cite{tao2021benchmarking, zhang2020machine, bejagam2022machine, ma2022machine, chen2020frequency, amrihesari2024machine} Early studies predominantly used feed-forward neural networks (fNNs) trained on curated polymer datasets, where polymers were encoded using handcrafted molecular descriptors or chemical fingerprints. As polymer databases expanded, researchers introduced convolutional neural networks (CNNs) that operate directly on 2D molecular images or one-hot encoded monomer sequences, capturing local substructural patterns that correlate with bulk properties such as modulus, toughness, and gas permeability. \cite{miccio2020chemical, park2022prediction}

NN approaches have also emerged as promising tools for predicting thermodynamic parameters and phase behavior in simplified polymer chain models by learning mappings from molecular representations and limited simulation or experimental datasets. Early demonstrations used supervised deep networks to infer $\chi$ parameters or solubility labels from simulation-generated morphologies or curated polymer–solvent datasets, showing that ML models can reproduce basic temperature-dependent trends and compress some chemistries into lower-dimensional, more intuitive predictors. \cite{tu2020machine, fang2025universal} More recent efforts emphasize physics-aware and multitask architectures that predict $\chi$ across many chemistries and temperatures, improving generalization by sharing representations across related tasks. These approaches can significantly accelerate mapping from chain composition to phase-relevant parameters compared with brute-force simulation. Hybrid strategies that combine theory (SCFT, Flory–Huggins) and simulations (CG, DPD) with NN models have also shown promise: embedding theoretical constraints while using simulations to generate synthetic training data reduces data requirements and improves extrapolation near phase boundaries. \cite{ethier2024predicting}

\begin{figure}[h!]
    \centering
    \includegraphics[width=0.66\linewidth]{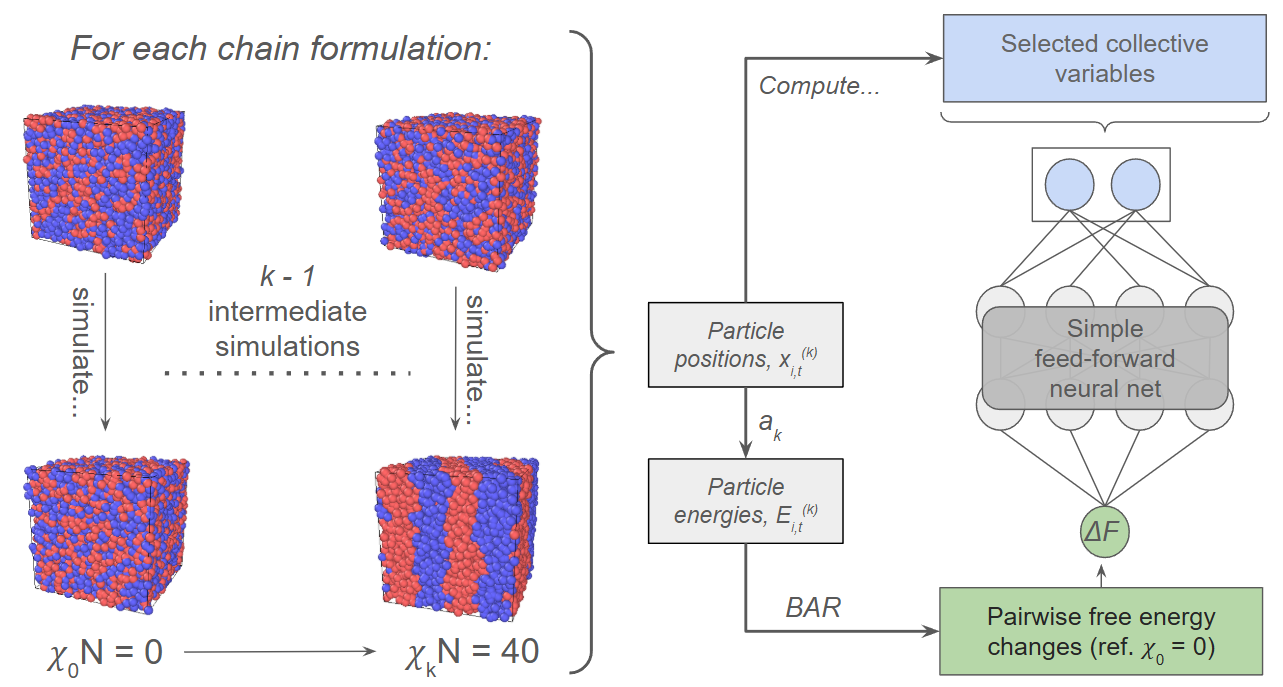}
    \caption{dBCP simulations are run across a spectrum of $\chi$ parameters, and a feed-forward neural network is trained on selected parameters to predict computed free energies.}
    \label{fig:para_opt}
\end{figure}

Motivated by these developments, this work aims to use ML to accelerate the prediction of polymer free energies. Free energies are fundamental thermodynamic quantities but are often expensive to compute, typically requiring extensive simulation and additional post-processing. By selecting appropriate collective variables (CVs) from simulations of linear BCP chains, we aim to train ML models capable of rapidly predicting BCP free energies without requiring additional simulation or computational cost. Utilizing ML to predict free energies may also provide a degree of model interpretability: by determining which variables the ML models prioritize during prediction, we can gain insight into the thermodynamic properties and behavior of these simulated BCP systems.

Every ML strategy requires the selection of suitable collective variables (CVs) to serve as inputs for neural networks that learn generalized mappings. As a starting point, we consider the simplest case: linear dBCPs with only two chemistries. Because these polymers are linear, monomer indexing—and therefore any per-bead quantities—has a natural and unambiguous ordering in vector form. This avoids complications that arise in nonlinear polymer architectures where multiple indexing conventions may exist. Focusing on this simpler system allows us to prioritize input parameter selection rather than challenges associated with molecular representation. It also enables the use of simple feed-forward neural networks (fNNs), which train rapidly and do not require more complex architectures. The general simulation–ML pipeline is outlined in Fig. \ref{fig:para_opt}. The metric we aim to model is the excess free energy of a dBCP relative to the corresponding homopolymer; that is, the free energy change between a dBCP with distinct chemistries in each block and a polymer with uniform chemistry. The free energies used as training targets are computed using the Bennett Acceptance Ratio (BAR). BAR is a Monte Carlo method for estimating free energies without the need for computationally expensive integral calculations by sampling per-chain energies between two states. \cite{bennett1976efficient} 

Although BAR is a state-of-the-art method for free energy computation, it struggles to accurately estimate free energies between states that differ significantly. For example, excess free energy is correlated with $\Delta \chi$. When $\Delta \chi$ is small, BAR can compute free energies with reasonable accuracy. However, for large $\Delta \chi$, or when states lie across a phase transition, BAR becomes unreliable due to insufficient phase-space overlap between sampled states. To mitigate this limitation, free energies are often computed using a \emph{stratified BAR} procedure, in which the free-energy difference is evaluated between a sequence of intermediate thermodynamic states and accumulated to obtain the total $\Delta F$. Because neighboring states are chosen to differ only slightly in $\chi N$, their configurational ensembles exhibit sufficient phase-space overlap, yielding reliable estimates. In this work we therefore treat the free energies obtained from the stratified BAR procedure as the reference (``true'') values. 

The breakdown of BAR when intermediate states are not used is illustrated in Fig.~\ref{fig:bar_overlap}, where direct BAR estimates closely track the reference free energies at small $\chi N$ but progressively deviate as $\chi N$ increases. In particular, the discrepancy becomes more pronounced in the strongly segregated regime, where the free energy differences between states grow and the configurational ensembles sampled at neighboring $\chi$ values share increasingly little overlap. As a consequence, the reweighting procedure underlying BAR becomes dominated by rare configurations and the resulting estimates exhibit systematic bias. This loss of overlap leads to increasing statistical error and bias in the BAR estimates, motivating the use of machine learning models to predict free energy differences in regimes where direct reweighting methods become unreliable.

Predicting free energy changes in dBCPs using neural networks would therefore both accelerate these calculations and potentially provide a stable approximation when BAR becomes inaccurate. Because BAR requires sampling from simulations to compute free energies, the input parameters used for ML prediction should reflect the statistical nature of these sampled configurations. One important consideration is the number of equilibrium samples used to compute both the free energies and the input parameters. Since the standard error of a sample mean scales as $1/\sqrt{N}$, where $N$ is the number of samples, selecting an adequate sample size without substantially increasing computational cost is important.

\begin{figure}[h!]
\centering
\begin{subfigure}{0.48\textwidth}
    \centering
    \caption{}
    \includegraphics[width=\linewidth]{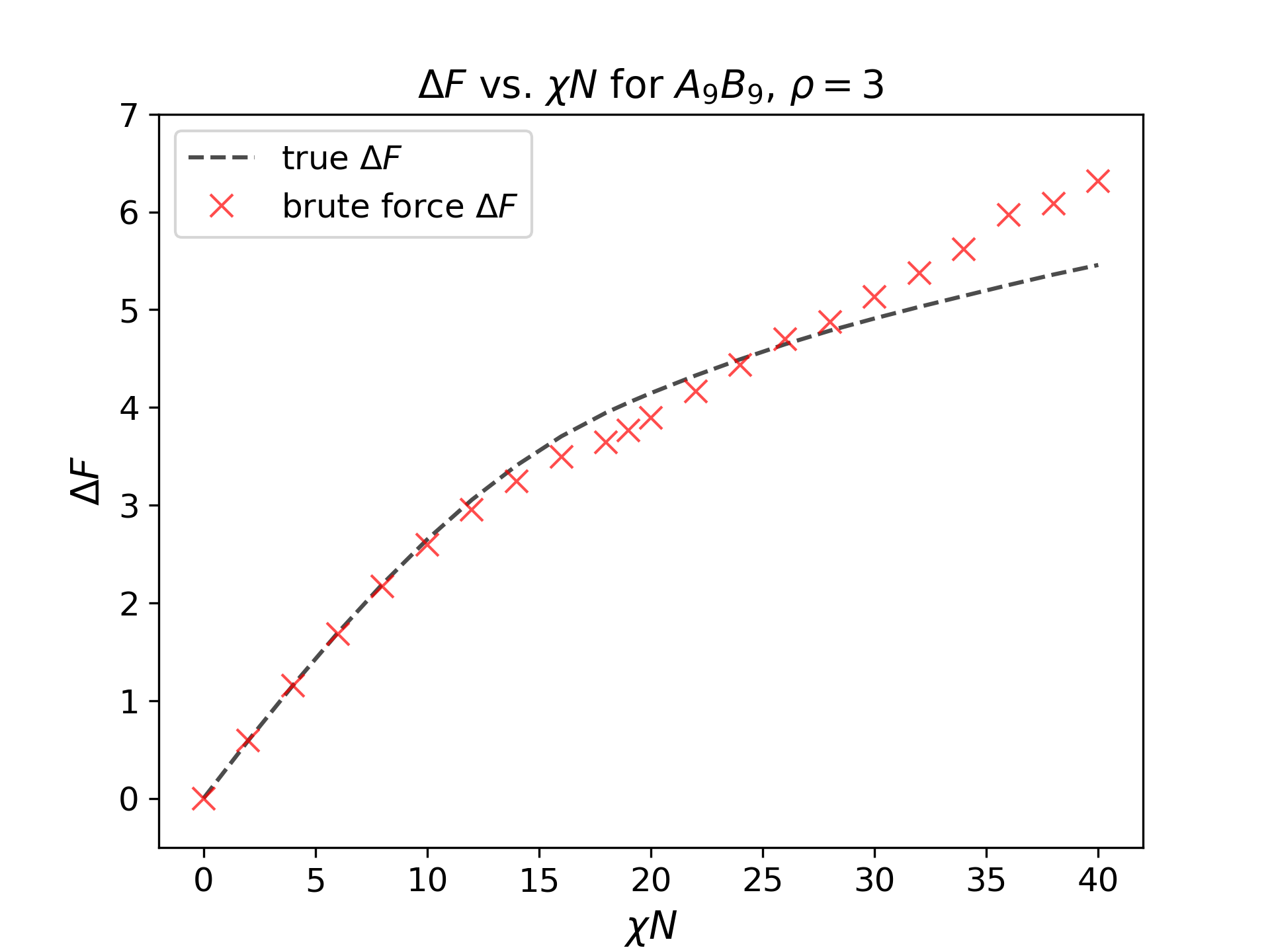}
    \label{fig:alpha_only}
\end{subfigure}
\begin{subfigure}{0.48\textwidth}
    \centering
    \caption{}
    \includegraphics[width=\linewidth]{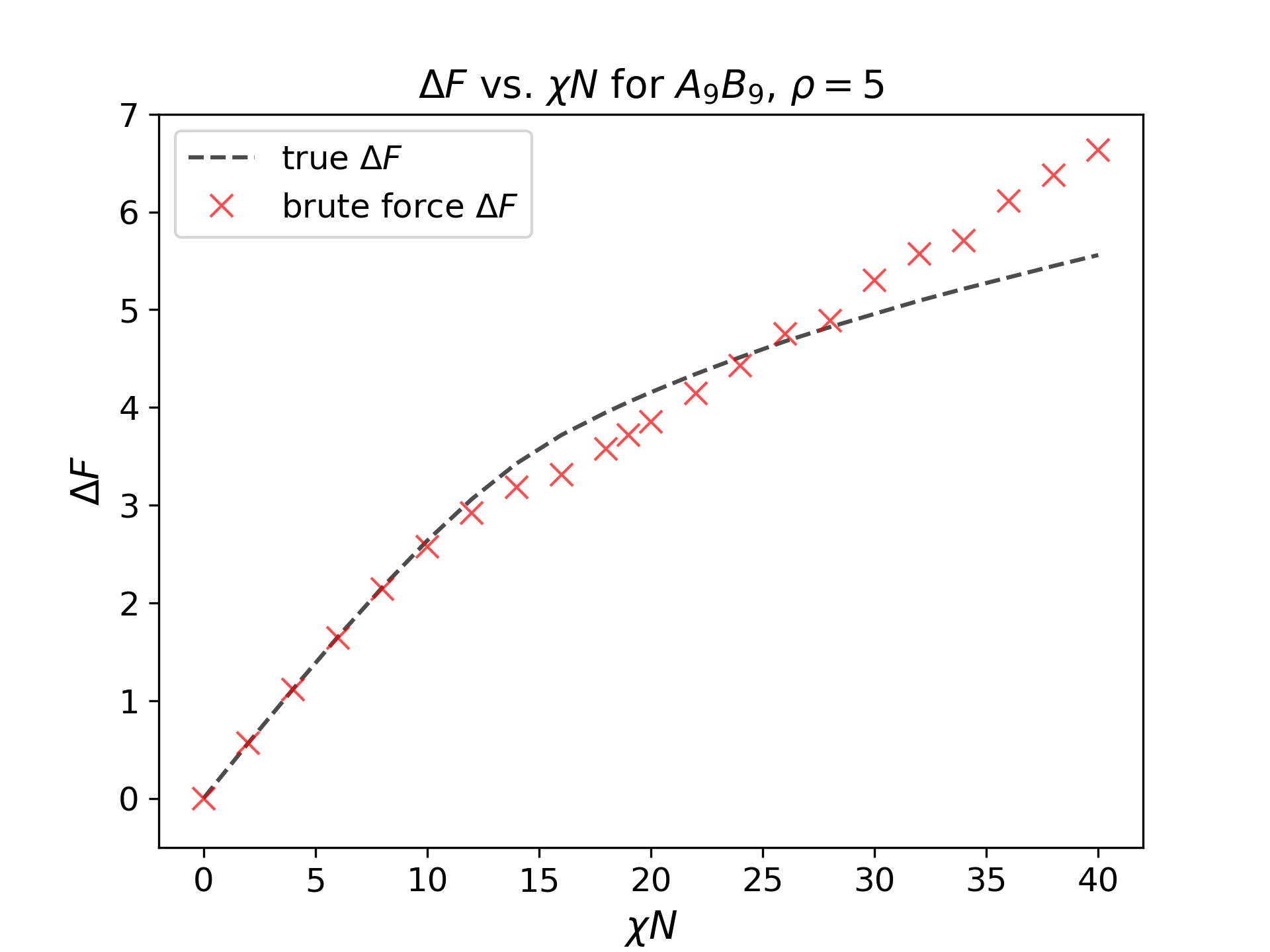}
    \label{fig:alpha_AB}
\end{subfigure}
\caption{Comparison between reference excess free energies $\Delta F$ (dashed line) and values estimated using the Bennett acceptance ratio (BAR) method (red crosses) for an $A_9B_9$ diblock copolymer as a function of $\chi N$. Results are shown for densities (a) $\rho=3$ and (b) $\rho=5$. While BAR accurately reproduces the free energy for small $\chi N$, deviations grow as $\chi N$ increases, reflecting the breakdown of phase-space overlap between sampled states.}
\label{fig:bar_overlap}
\end{figure}

For parameter selection, several candidate variables align naturally with existing theoretical intuition regarding polymer thermodynamics. Because free energies are fundamentally determined by the statistical distribution of microscopic states, energetic quantities provide a direct link between sampled configurations and the resulting thermodynamic observables. In particular, the potential energy contributions sampled during simulation reflect how chains interact with their environment and how those interactions evolve as the interaction parameter $\chi$ increases. 

Rather than using total per-chain energies, we consider per-bead energy contributions decomposed into heterogeneous interaction energies ($E_{AB}$), homogeneous interaction energies ($E_{\mathrm{hom}}$), and bonded spring energies ($E_K$). These quantities correspond directly to the physical interactions present in the coarse-grained model. The heterogeneous interaction energy captures unfavorable contacts between dissimilar beads and therefore reflects the driving force for microphase separation as $\chi$ increases. In contrast, homogeneous interaction energies characterize interactions within A- and B-rich domains and provide information about local packing and domain structure. The spring energy reflects deviations from the equilibrium bond length along the polymer backbone and therefore encodes the degree of chain stretching required to accommodate domain formation in the ordered phase.

To incorporate both equilibrium behavior and fluctuations in these energetic quantities, we compute the time-averaged mean and standard deviation of each per-bead energy contribution. These statistics summarize both the typical energetic environment experienced by each segment of the polymer chain and the extent to which that environment fluctuates over time. Because free energies depend not only on average energies but also on the distribution of accessible states, including both means and fluctuations provides a compact representation of the thermodynamic information contained within the sampled trajectories. In this sense, energetic descriptors provide a physically meaningful and computationally efficient representation of the microscopic states that determine polymer free energies.

Rapid and accurate free energy predictions would serve both as a benchmark and as a practical tool for designing inverse design algorithms. Free energy estimates allow comparison of stability between different dBCP states and can assist in determining whether a simulation has reached equilibrium, both of which could be performed much faster than with BAR calculations. Furthermore, since metastable states are common in dBCP phase separation, rapidly estimating free energies could help determine whether a simulation should be continued to reach the stable equilibrium state. Identifying relevant input parameters may also highlight key polymer properties important for inverse design and may aid in generalizing the approach to non-linear polymer architectures.


\section{Data selection and generation}
In order to construct a representative dataset of $AB$ linear block copolymers for model training, we select chain structures that span the parameter space defined by the total chain length $N$ and the fraction of $A$ monomers, $f_A$. Polymer chains are modeled using the freely-jointed chain (FJC) model with spring constant $K=100$ and equilibrium bond length $r_0=1$, a commonly employed coarse-grained representation in polymer simulation studies. The FJC model was chosen due to its relatively fast equilibration in dissipative particle dynamics (DPD) simulations, which serve as the primary simulation framework in this work. The set of chain structures considered is summarized in Table \ref{tab:chainset}. The entirety of the training set was utilized for model training and hyperparameter optimization. The holdout set was split in a 10\%/90\% for validation and testing, respectively.

\begin{table}[h!]
\centering
\renewcommand{\arraystretch}{1.3}
\begin{tabular}{c|c|c}
\hline
Dataset & Chain length ($N$) & Chain structures \\
\hline
\multirow{5}{*}{Training set} & $N = 10$   &      $A_5B_5$, $A_7B_3$      \\
& $N = 13$         &      $A_7B_6$, $A_{10}B_3$        \\
& $N = 15$        &         $A_7B_8$     \\
& $N = 17$        &      $A_9B_8$, $A_{13}B_4$        \\
& $N = 20$        &       $A_{10}B_{10}$, $A_{13}B_{7}$ , $A_{15}B_{5}$, $A_{17}B_{3}$        \\
\hline 
\multirow{4}{*}{\begin{tabular}{c}
Holdout set\\
(Validation / testing)
\end{tabular}} & $N = 12$   &      $A_6B_6$      \\
& $N = 14$        &     $A_7B_7$       \\
& $N = 16$        &      $A_8B_8$      \\
& $N = 18$        &     $A_9B_9$       \\
\hline
\end{tabular}
\vspace{2mm}
\caption{Selected chain structures for training and validation/testing.}
\label{tab:chainset}
\end{table}

Each chain structure was simulated using DPD within the NVT ensemble for $8\times10^6$ timesteps at two bead number densities, $\rho \in \{3,5\}$, which are commonly used parameterizations in coarse-grained polymer simulations. To sample a sufficiently broad $\chi N$ regime encompassing both disordered and ordered phases of the selected chain architectures, simulations were performed across $\chi N = \{0,2,4,\dots,40\}$. The Flory--Huggins interaction parameter $\chi$ was mapped to the DPD repulsion parameter $\alpha$ using the linear relation $\chi \mapsto z\alpha/kT$, where the coordination parameter $z \in \{0.284,\,0.661\}$ for densities $\rho=3$ and $\rho=5$, respectively, as reported by Petrov \textit{et al.}~\cite{petrov2024simple}. 

Because $\alpha$ is a direct simulation parameter in DPD, it is also included as an input feature to the machine learning models. The interaction parameter between identical beads was set to $a_{hom}=75/\rho$ to reproduce the compressibility of water, a standard choice in DPD simulations. The heterogeneous interaction parameter was then defined as
\[
a_{AB} = a_{hom} + \alpha .
\]

This symmetry induces a congruence between $A_xB_y$ and $A_yB_x$. Consequently, redundant simulations (e.g., $A_7B_3$ and $A_3B_7$) were avoided. During preprocessing, the per-bead feature vectors were reversed so that both orientations of the chain are represented, allowing the model to learn this structural equivalence during training.

To ensure equilibration prior to analysis, simulation data were recorded every 20{,}000 timesteps after the first $5\times10^6$ timesteps had elapsed, yielding a sufficiently large set of equilibrated configurations for subsequent calculations.

\medskip

Simulation output was processed in two stages: (1) computation of the per-bead potential energy components used as ML inputs, and (2) computation of free energies via BAR. For each stored configuration, we compute the per-bead homogeneous interaction energy ($E_{\text{hom}}$), heterogeneous interaction energy ($E_{AB}$), and spring energy ($E_K$). Homogeneous and heterogeneous energies correspond to interactions between identical ($A\!-\!A$ and $B\!-\!B$) and dissimilar ($A\!-\!B$) bead types, respectively, while the spring energy arises from bonded interactions along the polymer backbone.

Let $\Adj[i]$ denote the adjacency set of bead $i$, \footnote{Bead indices correspond to absolute indices in the simulation box instead of indices along an individual polymer chain to avoid confusion.} defined as all beads within a cutoff radius $r=1$ (including both bonded and nonbonded neighbors), and let $\textrm{B}[i]$ denote the set of beads bonded to bead $i$.%
\footnote{Note that $\textrm{B}[i] \not\subseteq \Adj[i]$ because bond stretching may extend beyond the cutoff radius $r=1$.}
We also define a bead-type function
\[
T(i): \mathbb{N} \rightarrow \{A,B\},
\]
which maps bead indices to their monomer type.

The per-bead energy contributions are then defined as
\[
E_{\text{hom}}(i) =
\frac{1}{4}
\sum_{j \in \Adj[i]}
\mathbbold{1}(T(i)=T(j)) \,
a_{hom} (r_{ij}-1)^2 ,
\]

\[
E_{AB}(i) =
\frac{1}{4}
\sum_{j \in \Adj[i]}
\mathbbold{1}(T(i)\neq T(j)) \,
a_{AB} (r_{ij}-1)^2 ,
\]

\[
E_K(i) =
\frac{1}{4}
\sum_{j \in \textrm{B}[i]}
K (r_{ij}-1)^2 ,
\]

where $r_{ij}$ denotes the Euclidean distance between beads $i$ and $j$.%

The total potential energy of bead $i$ is therefore
\[
E_{\text{tot}}(i) = E_{\text{hom}}(i) + E_{AB}(i) + E_K(i).
\]

For each bead $i$, the time-averaged mean and standard deviation of these energies across the sampled timesteps are computed as
\[
\expect{E(i)} =
\frac{1}{N_t}
\sum_t E^{(t)}(i),
\quad
\stdev{E(i)} =
\sqrt{
\frac{1}{N_t}
\sum_t
\left(E^{(t)}(i) - \expect{E(i)}\right)^2 } .
\]

We additionally define the bond covariance for a bonded pair $(i,j)$ with respect to an energy observable $E$ as
\[
\Cov{i}{j}
=
\expect{E(i)E(j)}
-
\expect{E(i)}\expect{E(j)} .
\]
%

For a polymer chain of length $N$, the vectors $\expect{E(i)}$ and $\stdev{E(i)}$ therefore have length $N$, while the covariance vector $\Cov{i}{j}$ has length $N-1$. The 10 primary input parameters tested during model training and optimization are shown in Table \ref{tab:input_para}. The selection and evaluation of optimal input feature combinations are discussed in the Results section.

\begin{table}[h!]
\centering
\renewcommand{\arraystretch}{1.3}
\begin{tabular}{c|c}
\hline
Quantity & Processed parameters  \\
\hline
$\alpha$ & $\alpha$\\
$E_{\text{hom}} $ & $\expect{\cdot}, \; \stdev{\cdot}, \; \Cov{\cdot}{\cdot}$\\
$E_{AB} $ & $\expect{\cdot}, \; \stdev{\cdot}, \; \Cov{\cdot}{\cdot}$\\
$E_{K} $ & $\expect{\cdot}, \; \stdev{\cdot}, \; \Cov{\cdot}{\cdot}$\\
\hline
\end{tabular}
\vspace{2mm}
\caption{Selected parameters and their processed equivalents for use in model inference.}
\label{tab:input_para}
\end{table}

Free-energy differences were computed using the Bennett Acceptance Ratio (BAR) method between states separated by $\Delta(\chi N)=2$, which ensures substantial configurational overlap between neighboring ensembles and therefore reliable estimates of $\Delta F$~\cite{bennett1976efficient}. The model is trained on free energies relative to the reference state $\chi N = 0$, since any constant free-energy offset can be absorbed into this reference value. To obtain free energies at larger $\chi N$, the incremental BAR estimates between adjacent states were summed sequentially from $\chi N = 0$ to the desired state. This stratified procedure significantly improves numerical stability compared with performing a single direct BAR calculation between the endpoint states.

To generate a sufficiently large dataset for model training, free energies were computed on a per-chain basis. Each simulation contains $1000$ polymer chains. Because reversing the bead ordering maps $A_xB_y$ chains onto equivalent $A_yB_x$ configurations, this symmetry effectively doubles the number of usable samples. As a result, each simulation yields $2000$ training data points rather than a single free-energy value for the entire system.

For each chain configuration sampled in the $\chi_i N=0$ ensemble, we evaluate the corresponding potential energies in both thermodynamic states. This produces four energy trajectories required for the BAR estimator: the sampled chain potential energies in the $\chi_i N$ and $\chi_f N$ simulations, along with the corresponding swapped-potential energies evaluated under the opposite Hamiltonian. Swapped energies are obtained by exchanging the potential functions between the two thermodynamic states and recomputing the energies. In the present case this operation is straightforward because the only parameter that changes with $\chi N$ is $a_{AB}$.

Let $M$ denote the set of beads belonging to a polymer chain. The total chain potential energy is

\[
E_{\text{\text{tot}}}
=
\sum_{i\in M} E_{\text{\text{tot}}}(i)
=
\sum_{i \in M}
\left(
E_{\text{hom}}(i)+E_{AB}(i)+E_K(i)
\right).
\]

Since only $E_{AB}$ depends on $\chi N$, the swapped potential energy can be written as

\[
E_{\text{swap}}
=
\sum_{i\in M}
\left(
E_{\text{hom}}(i)
+
\frac{a_{AB}^{\text{new}}}{a_{AB}^{\text{old}}}E_{AB}(i)
+
E_K(i)
\right).
\]

Given energy samples drawn from the two ensembles, BAR estimates the free-energy difference according to

\[
e^{-\beta (\Delta F + C)} =
\frac{
\left\langle
f\!\left(
\beta(E^{0\rightarrow\chi_f}_{\text{swap}}-E^{0}_{\text{\text{tot}}}-C)
\right)
\right\rangle_{0}
}{
\left\langle
f\!\left(
\beta(E^{\chi_f\rightarrow0}_{\text{swap}}-E^{\chi_f}_{\text{\text{tot}}}+C)
\right)
\right\rangle_{\chi_f}
},
\]

where $f(x)=1/(e^x+1)$ is the optimal BAR weighting function and $C \approx \Delta F$. The expectation values are taken over configurations sampled in the respective ensembles. This nonlinear equation is solved numerically to obtain $\Delta F$. In this work we use the \texttt{FastMBAR} Python package to efficiently solve the BAR equations across the large dataset generated from the simulations~\cite{ding2019fast}.

Once trained, the machine learning model eliminates the need to (1) numerically solve the BAR equations and (2) perform intermediate simulations at each $\Delta(\chi N)$ increment, both of which are substantially more computationally expensive than direct model inference.

\section{Model architecture and hyperparameter selection}
The model utilized in this work is a simple feed-forward neural network, implemented through the TensorFlow platform on the Keras API \cite{tensorflow2015-whitepaper}, as allowed by the linear nature of the simulated polymers. Because chains have variable lengths ($10 \leq N \leq 20$), energy-related feature vectors (see Table \ref{tab:input_para}) were padded to length 20 with a sentinel value of -1. These padding tokens were masked within the network to prevent them from contributing to the learned representations (the $\alpha$ parameter was not padded, as its dimension is constant with respect to $N$). As such, the input vector length, after concatenation of the selected input vectors, is fixed at $\dim(x)=20k + \mathbbold{1}(\alpha \textrm{ is included})$, where $k$ is the number of energy-related inputs selected. For example, if the model uses all $E_{AB}$-related inputs ($\mu, \sigma, \Cov{\cdot}{\cdot }$) and $\alpha$, $\dim(x) = 61$. The vector reversal introduced to account for congruent chains is applied before padding and concatenation, ensuring that reversed samples follow the same preprocessing pipeline as the original data. Experiments in determining optimal input parameters were initially performed with a 10 layer model, 512 nodes per layer; a batch size of 256; and mean squared error (MSE) as the primary loss function and evaluation metric. The Adam optimizer, with a initial learning rate of 0.001, was selected, and the model was trained for 40 epoches, if necessary, with an early stopping callback with respect to validation loss and a patience of 5. 

After optimal input parameters were determined with the model architecture outlined above, we perform a hyperparameter optimization on the hyperparameters outlined in Table \ref{tab:hpo_space}, with the \texttt{KerasTuner} Python package as the primary optimizer. \cite{omalley2019kerastuner}
\begin{table}[h!]
\centering
\renewcommand{\arraystretch}{1.3}
\begin{tabular}{c|c}
\hline
\textbf{Hyperparameter} & \textbf{Search Space} \\
\hline
Number of hidden layers ($L$) & $\{4,5,6,\dots,12\}$ \\

Nodes per layer ($N_{\mathrm{nodes}}$) & $\{64, 96, 128, \dots, 1024\}$ \\

Activation function & $\{\mathrm{ReLU},\ \mathrm{LeakyReLU}(\alpha)\}$ \\

LeakyReLU slope ($\alpha$) & $\{0.01,\ 0.1,\ 0.2,\ 0.3\}$ \\

Learning rate ($\eta$) & $\{10^{-1},\ 10^{-2},\ 10^{-3},\ 10^{-4}\}$ \\

Training loss function & $\{\mathrm{MSE},\ \mathrm{MAE},\ \mathrm{Huber}(\delta=1),\ \mathrm{Huber}(\delta=0.5)\}$ \\

Batch size ($B$) & $\{32,\ 64,\ 128,\ 256,\ 512\}$ \\

\hline
\end{tabular}
\vspace{2mm}
\caption{Hyperparameters optimized during model selection and their corresponding search spaces. Model configurations were evaluated using validation mean squared error (MSE) as the optimization objective.}
\label{tab:hpo_space}
\end{table}

Validation MSE was the primary evaluation metric, and the optimization was run for 40 trials across the selected hyperparameters. The input selection experiments were then performed with these optimized hyperparameters to ensure no significant changes in model performance with respect to input.

\section{Results and discussion}

\subsection{Baseline models}

As a baseline for evaluating the predictive value of the proposed input parameters, we first train models using only the interaction parameter $\alpha$, followed by a model incorporating both $\alpha$ and the average heterogeneous interaction energy $\expect{E_{AB}}$. The parameter $\alpha$ is directly proportional to the Flory--Huggins interaction parameter $\chi$, and therefore to $\chi N$. Because the excess free energy $\Delta F$ is computed relative to the reference state $\chi N = 0$, one expects a monotonic relationship between $\alpha$ and $\Delta F$. In the weak-segregation regime (small $\chi N$), theoretical treatments such as self-consistent field theory predict an approximately linear scaling $\Delta F \propto \chi N$. However, at larger $\chi N$ values the system undergoes microphase separation and the free energy becomes dominated by entropic penalties associated with chain stretching and interface formation \cite{leibler1980theory}. Consequently, the relationship between $\alpha$ and $\Delta F$ becomes strongly nonlinear in this regime.

\begin{figure}[h!]
\centering
\begin{subfigure}{0.45\textwidth}
    \centering
    \caption{}
    \includegraphics[width=\linewidth]{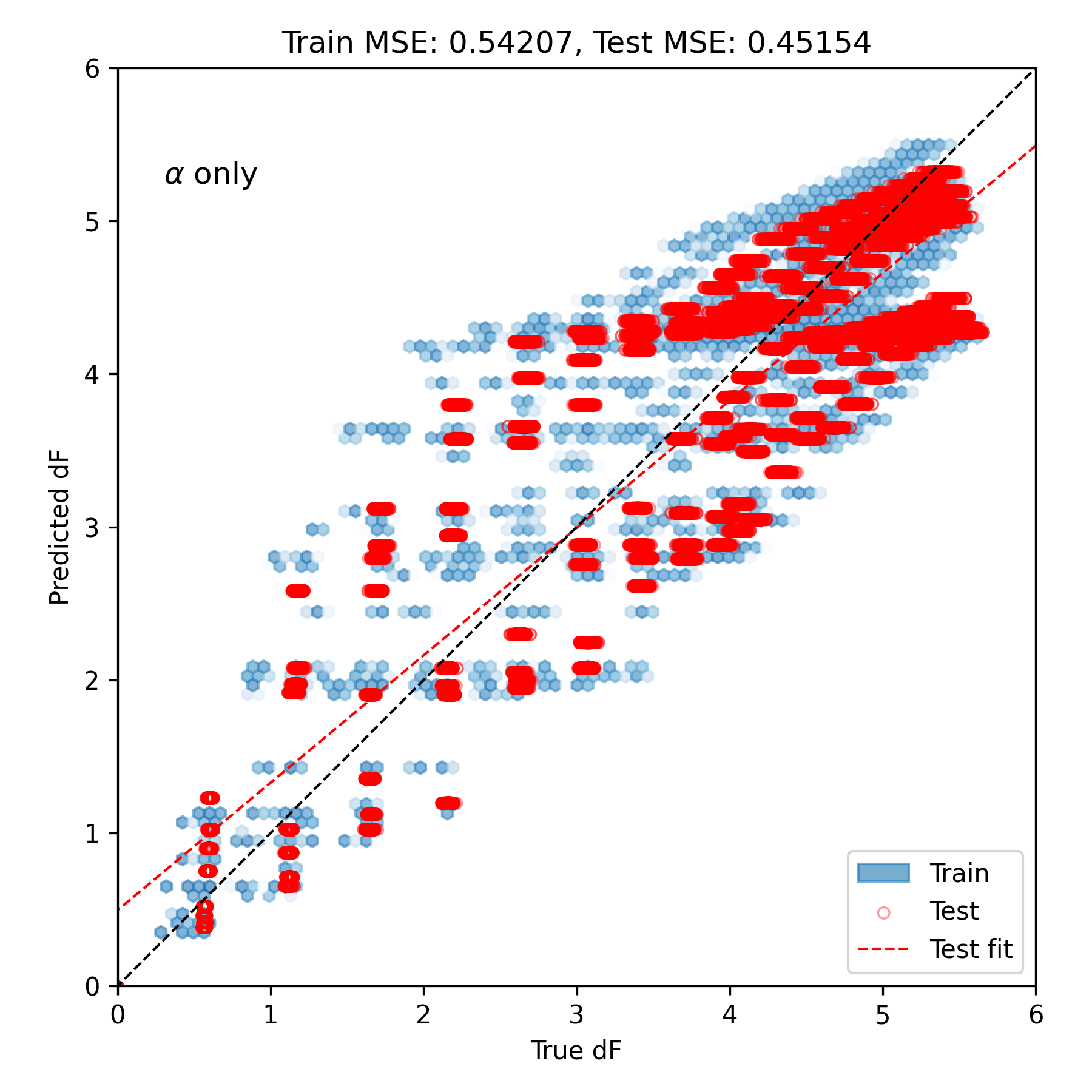}
    \label{fig:alpha_only}
\end{subfigure}
\begin{subfigure}{0.45\textwidth}
    \centering
    \caption{}
    \includegraphics[width=\linewidth]{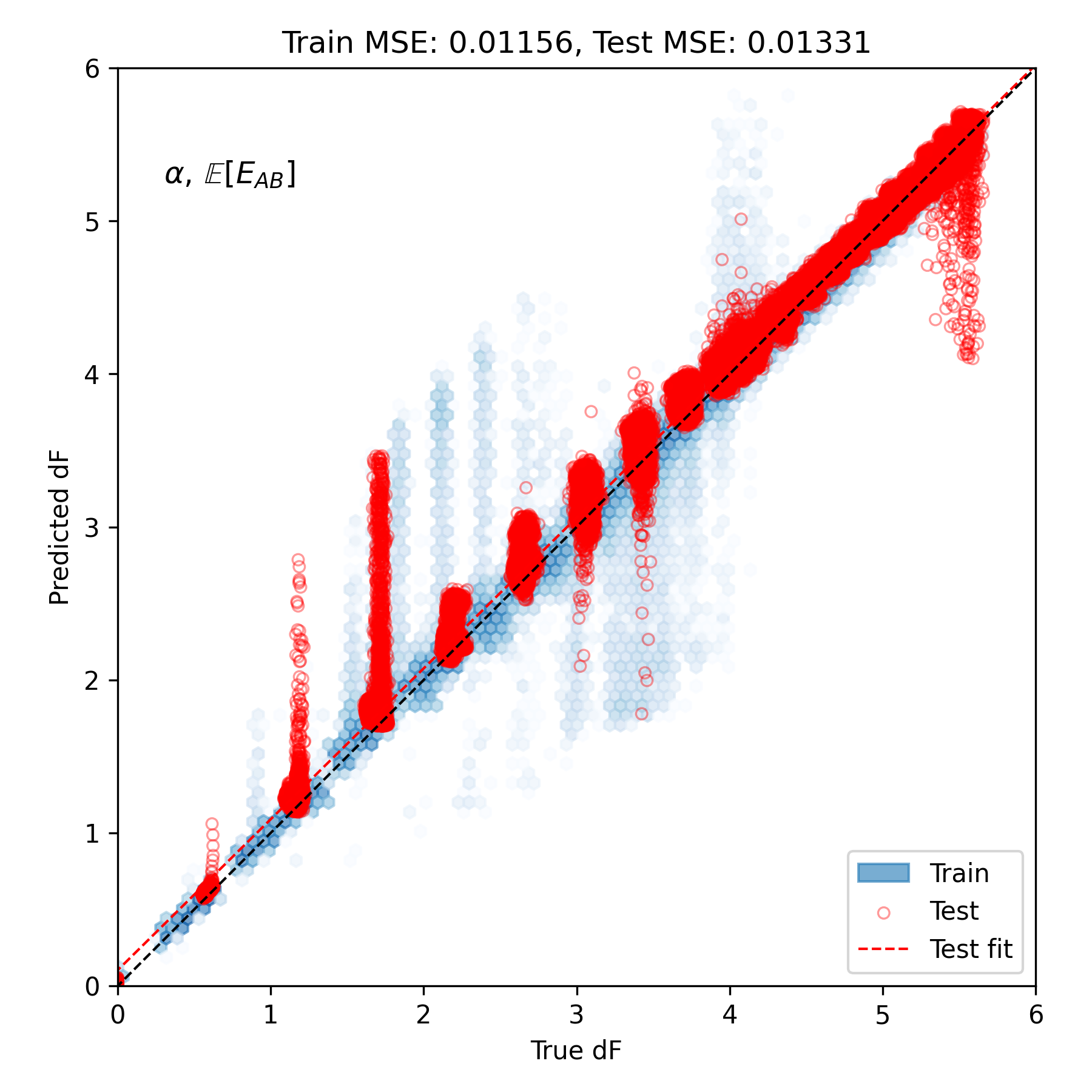}
    \label{fig:alpha_AB}
\end{subfigure}

\caption{Training and testing output parity plots for \textbf{(a)} the $\alpha$ only model and \textbf{(b)} the $\alpha$, $\expect{E_{AB}}$ model. The black $y=x$ line represents perfect parity. The discrepancy between training and testing MSE is a consequence of the restricted testing dataset, which is not observed in later models.}
\label{fig:baseline}
\end{figure}

The parity plots for the training and testing datasets for the $\alpha$-only model are shown in Fig.~\ref{fig:baseline}(a). As expected, $\alpha$ alone is insufficient to accurately predict free energies. Although the model captures the trivial positive correlation between $\alpha$ and $\Delta F$, the predictions exhibit large scatter across the full range of outputs. This result is consistent with the fact that $\alpha$ encodes only the interaction strength and contains no information about chain architecture or the spatial distribution of $A-B$ contacts within the polymer.

To incorporate structural information, we next introduce the mean heterogeneous interaction energy $\expect{E_{AB}}$ as an additional input feature. This vector implicitly encodes two important aspects of the polymer system. First, the length of the vector reflects the chain length $N$, allowing the model to distinguish polymers with different architectures. Second, the spatial variation of $E_{AB}$ along the chain captures how A--B interactions change with increasing $\alpha$, providing information about the degree of segregation and the location of interfacial regions.

The resulting parity plots are shown in Fig.~\ref{fig:baseline}(b). Incorporating $\expect{E_{AB}}$ leads to a substantial improvement in predictive accuracy, reducing the average testing error by approximately a factor of seven. Despite this improvement, the model still exhibits significant uncertainty across portions of the free-energy range. In particular, predictions around $\Delta F \approx 0.5$ display large variability, with predicted values spanning nearly the entire range between 0 and 1. Additionally, the model tends to systematically overestimate free energies in the weak-segregation regime. For example, chains with $\Delta F \approx 1.8$ are sometimes predicted to have free energies approaching $\Delta F \approx 3.5$.

Interestingly, the order--disorder transition (ODT) becomes visually apparent in these parity plots. Predictions corresponding to the disordered regime rarely exceed $\Delta F \approx 4$, whereas values in the ordered regime tend to cluster above this threshold. This suggests that the inclusion of $\expect{E_{AB}}$ allows the model to infer phase behavior from interaction energy patterns along the chain. Nevertheless, the baseline model still struggles to accurately predict free energies for chains that share similar phases but differ in composition or chain length, motivating the introduction of additional energetic descriptors.

\subsection{Incorporating different energy terms}

The total potential energy of a bead consists of three contributions: the heterogeneous interaction energy $E_{AB}$, the homogeneous interaction energy $E_{\text{hom}}$, and the bonded spring energy $E_K$. While the heterogeneous interaction energy directly reflects unfavorable A--B contacts, the homogeneous and spring energy components capture additional information about local packing and chain conformations. Incorporating these quantities may therefore allow the model to distinguish between polymer chains that exhibit similar A--B interaction patterns but differ in internal conformational structure.

\begin{figure}[h!]
\centering
\begin{subfigure}{0.45\textwidth}
    \centering
    \caption{}
    \includegraphics[width=\linewidth]{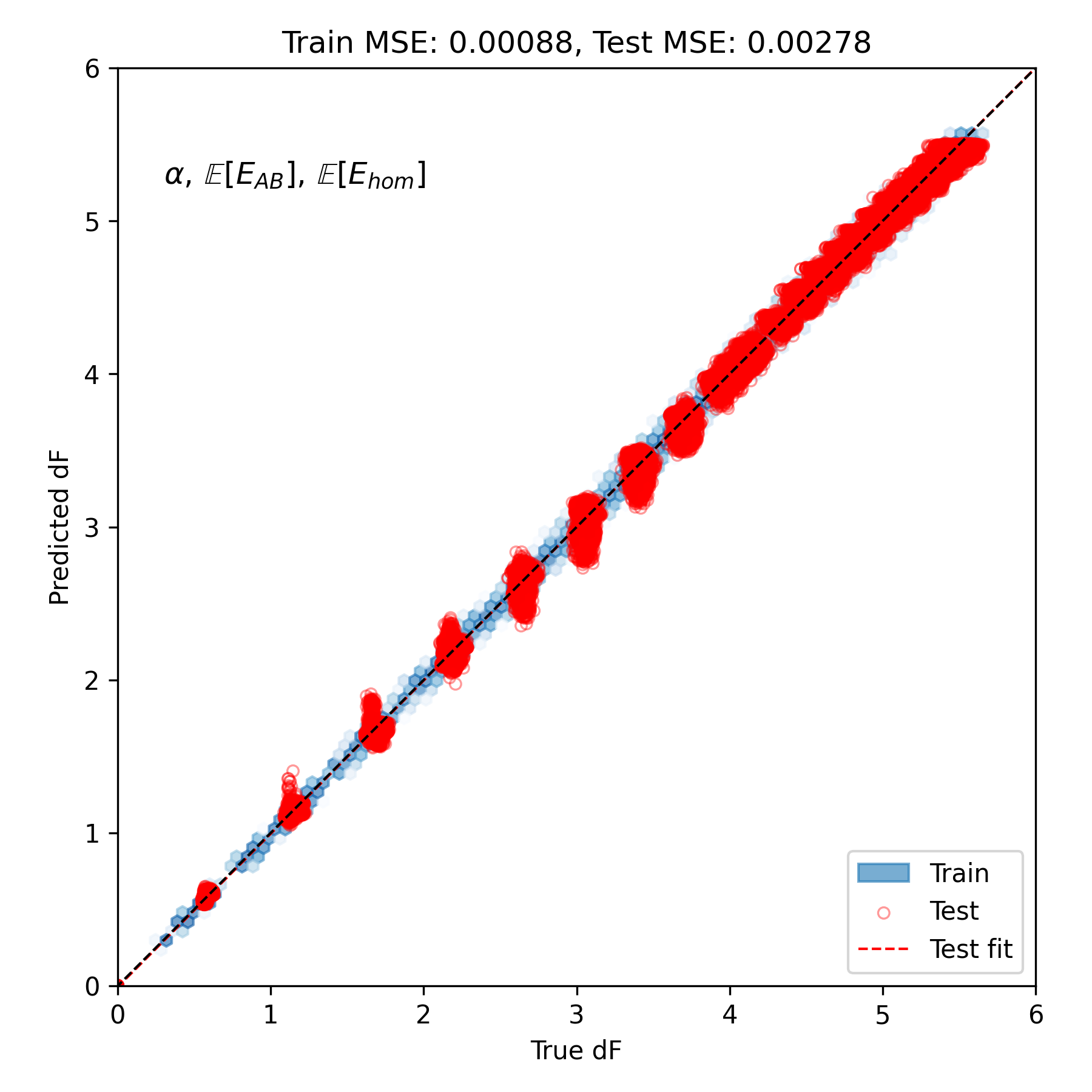}
    \label{fig:alpha_only}
\end{subfigure}
\begin{subfigure}{0.45\textwidth}
    \centering
    \caption{}
    \includegraphics[width=\linewidth]{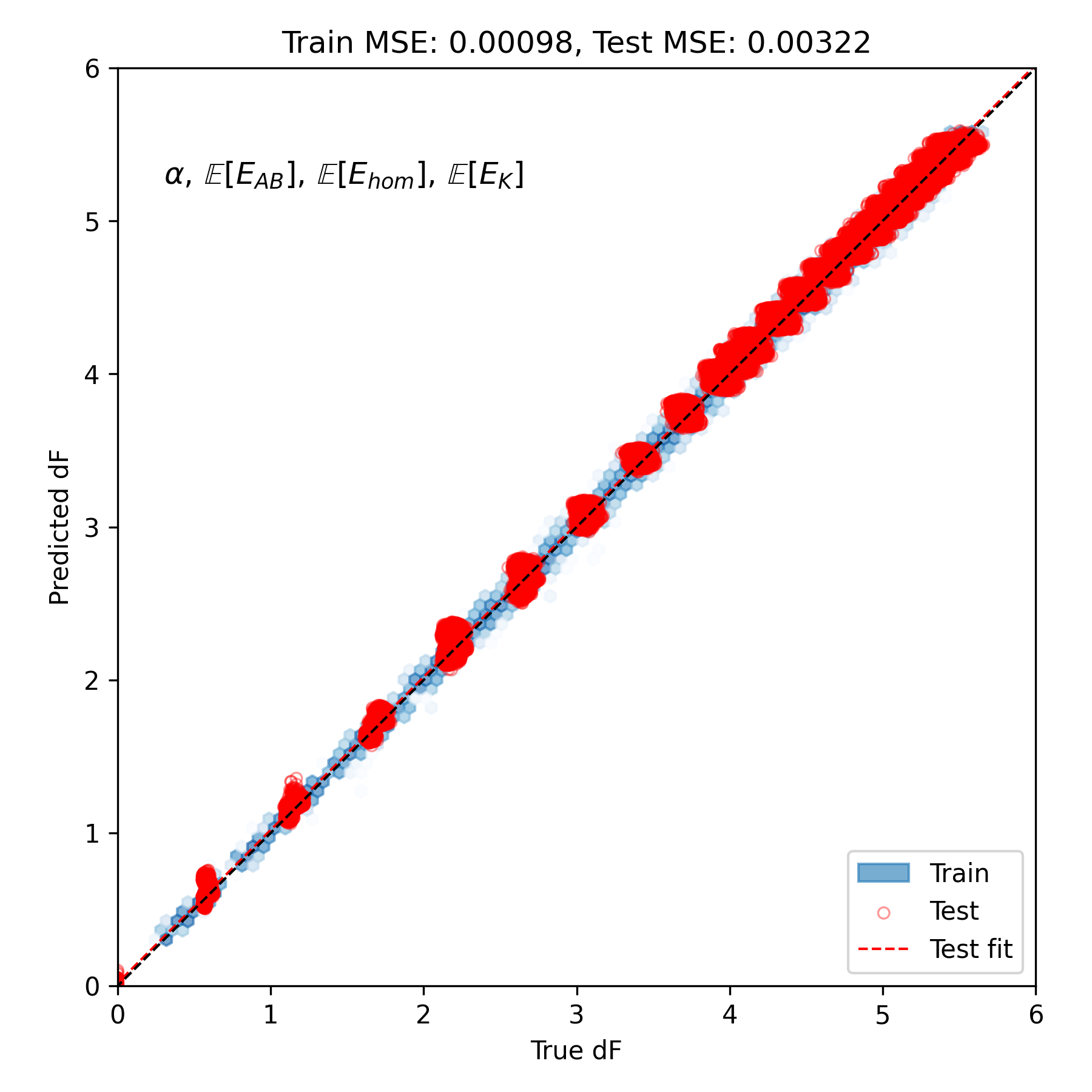}
    \label{fig:alpha_AB}
\end{subfigure}

\caption{Training and testing output parity plots for \textbf{(a)} the $\alpha$, $\expect{E_{AB}}$, $\expect{E_{\text{hom}}}$ model and \textbf{(b)} the $\alpha$, $\expect{E_{AB}}$, $\expect{E_{\text{hom}}}$, $\expect{E_{K}}$ model. The black $y=x$ line represents perfect parity.}
\label{fig:energy_terms}
\end{figure}

We first augment the baseline model by including the mean homogeneous interaction energy $\expect{E_{\text{hom}}}$ together with $\alpha$ and $\expect{E_{AB}}$. As shown in Fig.~\ref{fig:energy_terms}(a), the resulting parity plot demonstrates a substantial improvement in both training and testing accuracy compared to the baseline models. The predicted free energies closely follow the parity line across the entire output range, indicating that the additional energetic information significantly reduces ambiguity in the mapping between interaction parameters and free energy.

The improvement can be understood by considering the physical interpretation of $E_{\text{hom}}$. While $E_{AB}$ primarily measures interfacial contacts between dissimilar beads, $E_{\text{hom}}$ captures interactions within the A and B domains themselves. These contributions provide information about local packing density and domain structure, which are directly related to the entropic penalties associated with chain stretching in the ordered phase.

We further extend the model by incorporating the mean spring energy $\expect{E_{K}}$, which reflects bond stretching along the polymer backbone. Because the freely-jointed chain model penalizes deviations from the equilibrium bond length, the spring energy provides a direct measure of conformational distortion within the chain. In the strongly segregated regime, chains must stretch in order to maintain domain separation, making this contribution particularly relevant for describing high-$\chi N$ states.

The parity plot for the model including all three mean energy terms is shown in Fig.~\ref{fig:energy_terms}(b). For the present dataset, incorporating $\expect{E_{K}}$ yields only a marginal change in predictive accuracy relative to the model that excludes this term. This result is not unexpected given the specific simulation conditions used in this study. The freely-jointed chain model employs a relatively stiff harmonic spring potential with fixed parameters ($K=100$, $r_0=1$), meaning that bond-length fluctuations remain small and contribute only modestly to variations in the total free energy across the sampled configurations.

Nevertheless, including the spring energy remains important for the generality of the approach. In more flexible chain models—such as Gaussian chain representations or simulations with varying spring constants—bond stretching can play a much larger role in determining chain conformations and free energies. In experimental systems as well, polymer chains may experience substantial stretching under confinement or during microphase separation, leading to stronger coupling between conformational entropy and interaction energies. Under such conditions, the spring-energy contribution may become a critical descriptor for accurately predicting free energies. Incorporating $\expect{E_{K}}$ therefore provides a more physically complete representation of the chain energetics and ensures that the machine learning framework can readily extend to polymer models in which chain stretching is a dominant thermodynamic factor.

\subsection{Incorporating variances and correlations}

The models discussed above rely solely on time-averaged energy quantities. However, free energies are fundamentally related to the statistical distribution of microscopic configurations rather than only their mean properties. To incorporate information about fluctuations in chain conformations and interactions, we next include the standard deviations of the per-bead energy components.

\begin{figure}[h!]
\centering
\begin{subfigure}{0.45\textwidth}
    \centering
    \caption{}
    \includegraphics[width=\linewidth]{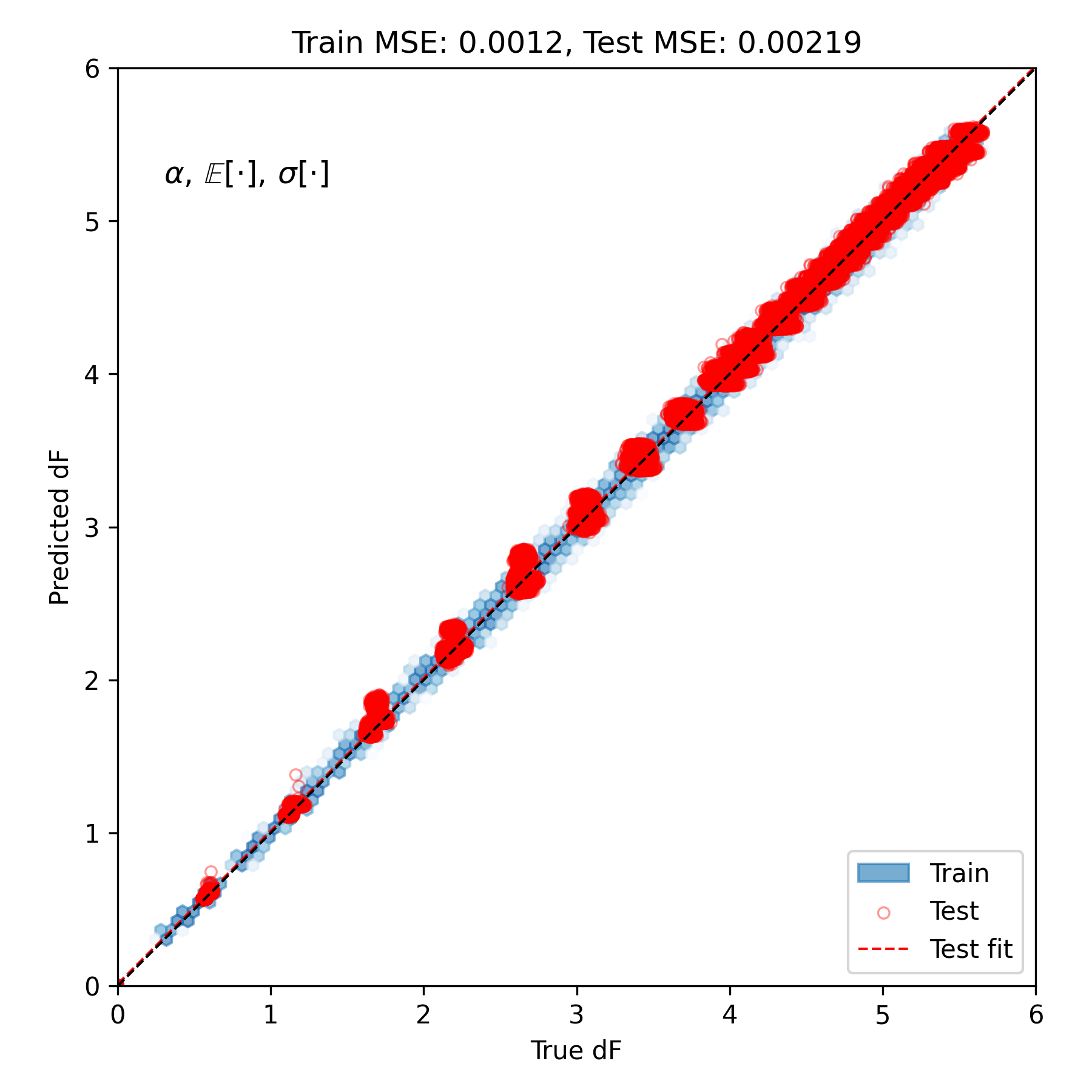}
    \label{fig:alpha_only}
\end{subfigure}
\begin{subfigure}{0.45\textwidth}
    \centering
    \caption{}
    \includegraphics[width=\linewidth]{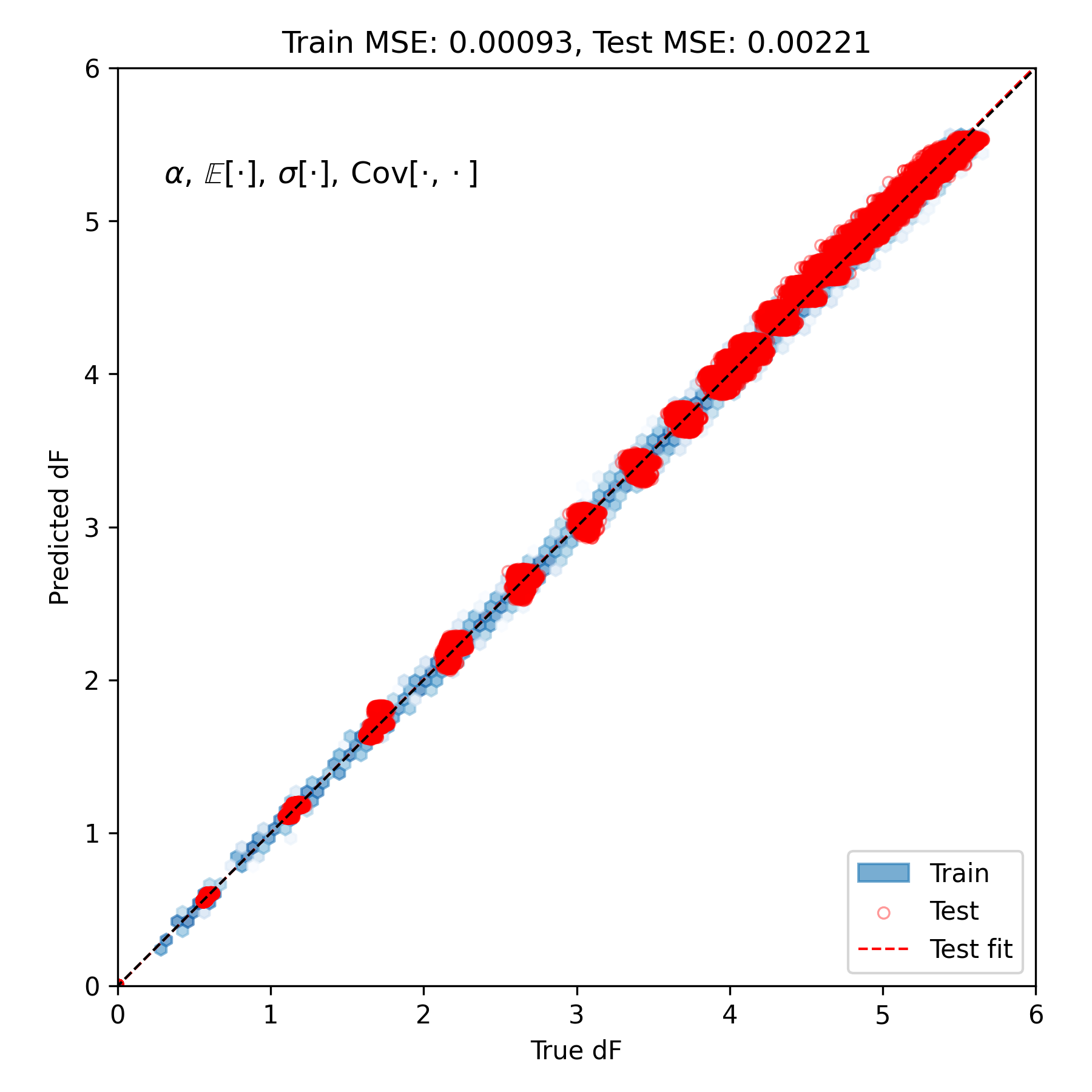}
    \label{fig:alpha_AB}
\end{subfigure}
\caption{Training and testing output parity plots for \textbf{(a)} the $\alpha$, $\expect{\cdot}$, $\stdev{\cdot}$ model and \textbf{(b)} the $\alpha$, $\expect{\cdot}$, $\stdev{\cdot}$, $\Cov{\cdot}{\cdot}$ model, both of which incorporate all three energy terms outlined previously. The black $y=x$ line represents perfect parity.}
\label{fig:variances}
\end{figure}

Specifically, we train a model using $\alpha$ together with both the mean and standard deviation of each energy contribution, denoted $\alpha, \, \expect{\cdot}, \, \sigma[\cdot]$. The resulting parity plot is shown in Fig.~\ref{fig:variances}(a). Including these variance terms slightly improves predictive performance relative to the models that rely only on mean energies. The improvement suggests that fluctuations in local interaction energies provide additional information about the accessible configuration space of the polymer chain. Physically, these fluctuations reflect the degree of conformational freedom available to the chain, which directly influences the entropic contribution to the free energy.

To further incorporate correlations between neighboring beads, we introduce the covariance of bonded pairs as an additional input feature. These covariance terms quantify the extent to which fluctuations in the energies of adjacent beads are correlated, thereby capturing cooperative effects along the polymer backbone. The corresponding model, which includes $\alpha$, $\expect{\cdot}$, $\sigma[\cdot]$, and $\mathrm{Cov}[\cdot,\cdot]$, is shown in Fig.~\ref{fig:variances}(b).

The inclusion of covariance terms yields only a marginal improvement in predictive accuracy compared to the variance-only model. This result suggests that most of the relevant thermodynamic information is already captured by the first and second moments of the individual energy distributions. While covariance terms introduce additional structural information about correlations along the chain, their contribution to predicting equilibrium free energies appears to be comparatively small for the present system.

Nevertheless, covariance features may become more important in situations involving nonequilibrium dynamics or driven polymer systems. In such scenarios, correlations between neighboring segments can influence how energy fluctuations propagate along the chain, potentially affecting the work distributions that determine free-energy differences through relations such as Jarzynski equality or Crooks fluctuation theorem. \cite{jarzynski1997nonequilibrium, crooks1998nonequilibrium} As a result, correlated energy fluctuations may play a larger role in predicting effective free energies in nonequilibrium conditions.

For these reasons, covariance descriptors are retained in the broader feature set used during model selection and hyperparameter optimization. Including these additional descriptors allows the learning architecture to adapt to richer feature spaces and ensures that the resulting model remains robust if more complex polymer models or nonequilibrium simulation data are incorporated in future studies. After hyperparameter optimization identifies an appropriate network architecture, feature importance can then be reassessed to determine the minimal set of descriptors required for accurate predictions.

\subsection{Optimizing and verifying hyperparameters}
\begin{table}[h!]
\centering
\renewcommand{\arraystretch}{1.3}
\begin{tabular}{c|c}
\hline
\textbf{Hyperparameter} & \textbf{Optimal parameter} \\
\hline
Number of hidden layers ($L$) & 5 \\

Nodes per layer ($N_{\mathrm{nodes}}$) & 288 \\

Activation function & $\mathrm{LeakyReLU}(\alpha = 0.2)$ \\

Learning rate ($\eta$) & $10^{-3}$ \\

Training loss function & $\mathrm{Huber}(\delta=1)$ \\

Batch size ($B$) & 32 \\

\hline
\end{tabular}
\vspace{2mm}
\caption{Optimal hyperparameters obtained after 40 trials using the \texttt{KerasTuner} package.}
\label{tab:hpo_opt}
\end{table}

After identifying the most informative input parameters, we performed a hyperparameter optimization over the network architecture and training parameters using the search space summarized in Table~\ref{tab:hpo_space}. The optimal configuration obtained after 40 trials is shown in Table~\ref{tab:hpo_opt}.

To evaluate the effect of these optimized hyperparameters, we retrained the models from Sections~4.1--4.3 using the optimized architecture and compared their test mean squared errors (MSE) to those obtained using the initial architecture ($10$ hidden layers with $512$ nodes per layer). The comparison is summarized in Table~\ref{tab:base_vs_hpo}. In general, hyperparameter optimization did not produce a uniform improvement across all model variants. For the baseline and intermediate models, the optimized architecture yielded similar or slightly worse performance relative to the original architecture. In particular, the model using $\alpha$ and $\expect{E_{AB}}$ exhibited an increase in test error after optimization, suggesting that the larger baseline network already provided sufficient capacity to capture the relationship encoded by these descriptors.

\begin{table}[h!]
\centering
\renewcommand{\arraystretch}{1.3}
\begin{tabular}{c|c|c|c}
\hline
Category & Model & Initial test MSE & Optimized test MSE \\
\hline
\multirow{2}{*}{Baseline models (4.1)} & $\alpha$ only & 0.452 &  0.435  \\
 & $\alpha$, $\expect{E_{AB}}$ & 0.0133 &   0.0234  \\
 \hline
 \multirow{2}{*}{Energy models (4.2)}  &$\alpha$, $\expect{E_{AB}}$, $\expect{E_{H}}$ & 0.00278 &   0.00206  \\
 & $\alpha$, $\expect{E_{AB}}$, $\expect{E_{H}}$, $\expect{E_{K}}$ & 0.00322 &   0.00495  \\
  \hline
  \multirow{2}{*}{Energy-variance models (4.3)}  & $\alpha$, $\expect{\cdot}$, $\stdev{\cdot}$ & 0.00219 &   0.00387   \\
   & $\alpha$, $\expect{\cdot}$, $\stdev{\cdot}$, $\Cov{\cdot}{\cdot}$ & 0.00221 &   0.00190  \\

\hline
\end{tabular}
\vspace{2mm}
\caption{Test MSEs between the inital model architecture vs. the hyperparameter-optimized model architecture.}
\label{tab:base_vs_hpo}
\end{table}

\begin{figure}[h!]
\centering
\begin{subfigure}{0.45\textwidth}
    \centering
    \caption{}
    \includegraphics[width=\linewidth]{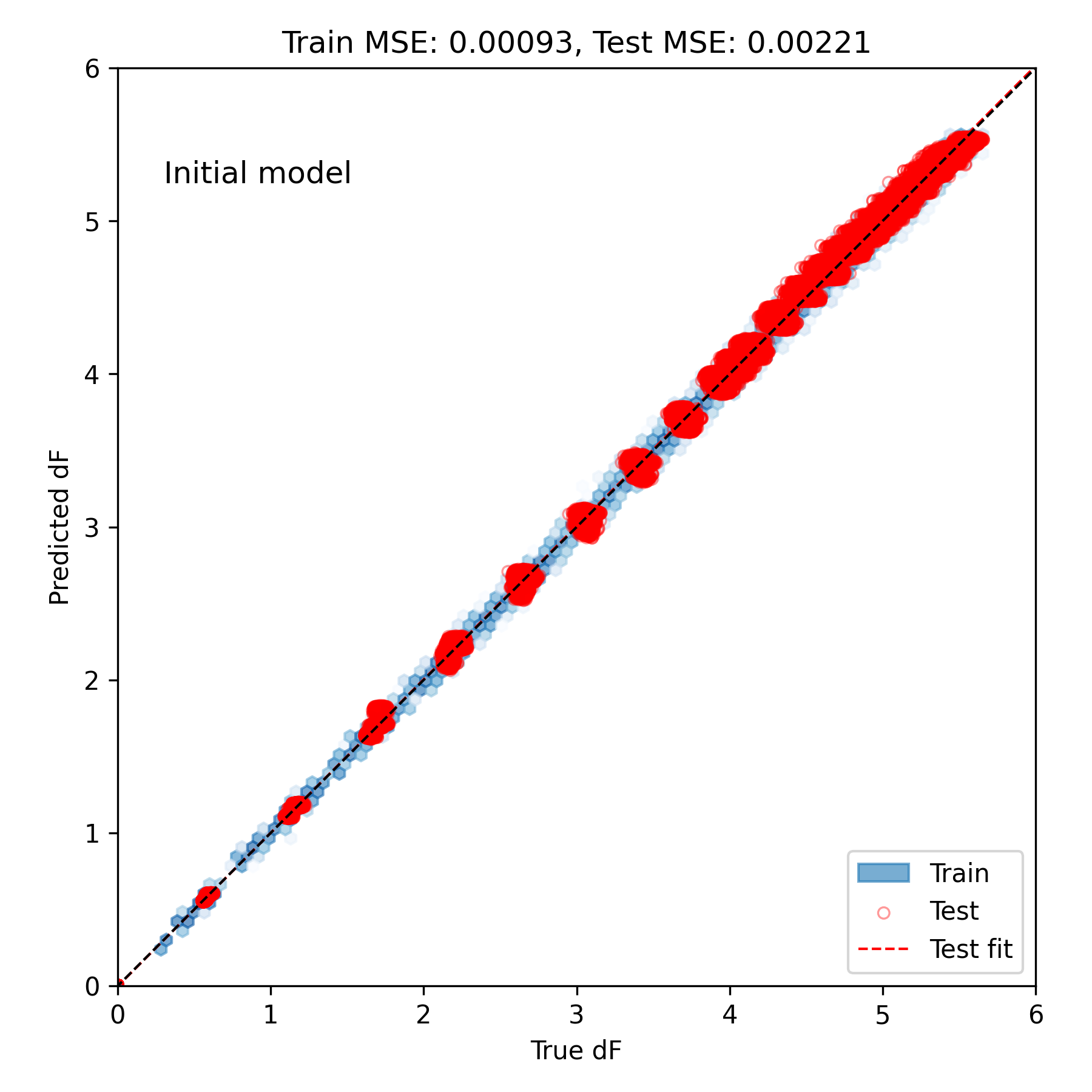}
    \label{fig:alpha_only}
\end{subfigure}
\begin{subfigure}{0.45\textwidth}
    \centering
    \caption{}
    \includegraphics[width=\linewidth]{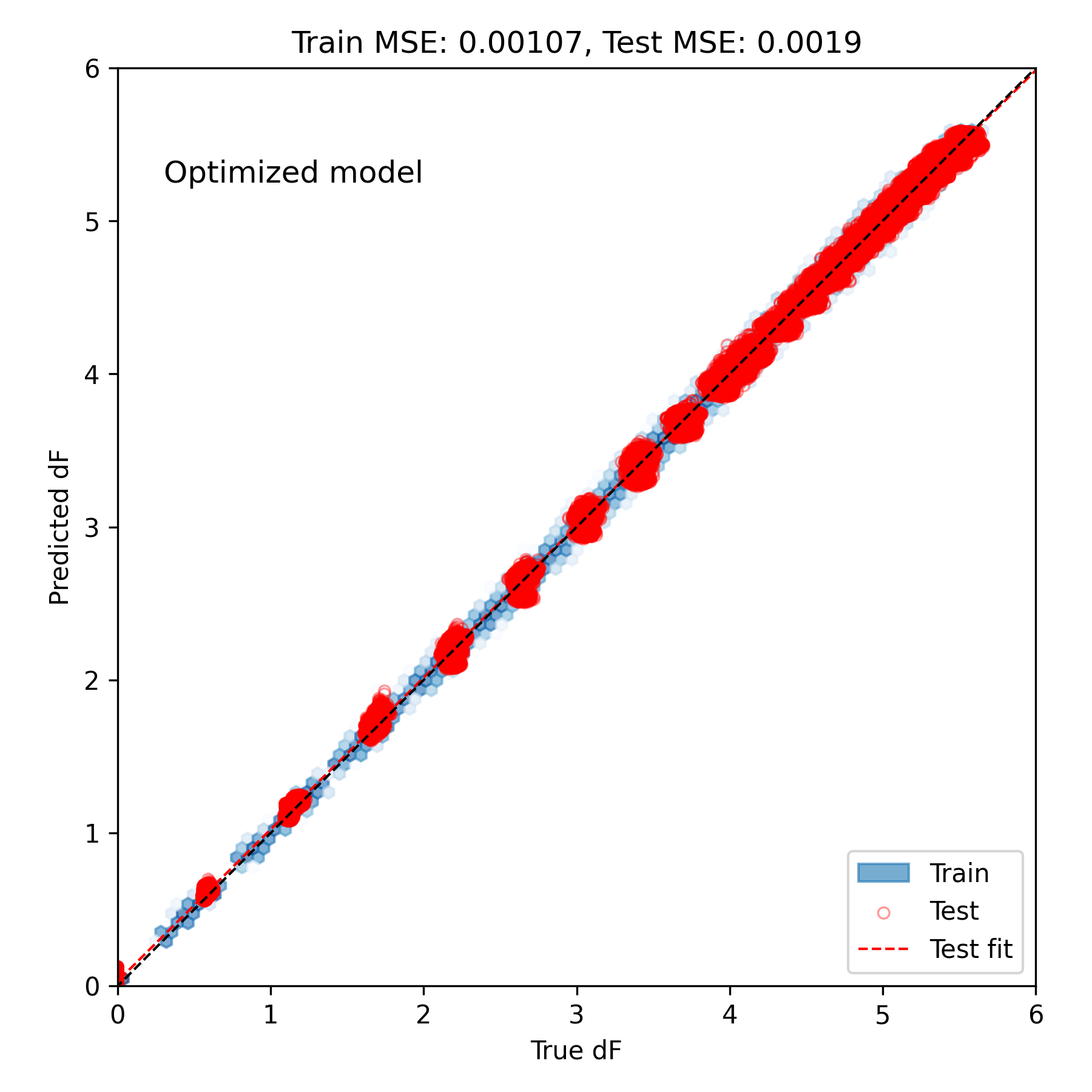}
    \label{fig:alpha_AB}
\end{subfigure}

\caption{Training and testing output parity plots for \textbf{(a)} the initial experimental model and \textbf{(b)} the hyperparameter optimized $\alpha$, $\expect{\cdot}$, $\stdev{\cdot}$, $\Cov{\cdot}{\cdot}$ model, outlined previously. The black $y=x$ line represents perfect parity.}
\label{fig:base_vs_hpo_plot}
\end{figure}

In contrast, the most expressive model incorporating the full set of energetic statistics $(\alpha, \expect{\cdot}, \sigma[\cdot], \Cov{\cdot}{\cdot})$ showed a modest improvement after hyperparameter optimization, achieving the lowest test MSE among all models considered (Fig.~\ref{fig:base_vs_hpo_plot}. This improvement is consistent with the fact that the hyperparameter search was performed using this most expressive feature set; consequently, the selected architecture is likely better suited to modeling the higher-dimensional feature space associated with these descriptors.

The intermediate energy model incorporating $\expect{E_{AB}}$ and $\expect{E_{H}}$ displayed performance comparable to the model that also includes $\expect{E_{K}}$. As discussed previously, the limited improvement obtained by including the spring-energy contribution may arise from the specific conditions of the present dataset. The freely-jointed chain model employs a relatively stiff harmonic spring potential, which restricts bond-length fluctuations and reduces the variability of $E_K$ across sampled configurations. In addition, the training dataset contains relatively few chain architectures that strongly emphasize conformational stretching effects. As a result, the contribution of $E_K$ to the predictive mapping between energetic descriptors and free energy remains small in the current dataset. Expanding the dataset to include additional chain architectures with stronger conformational variation, or simulations that more fully sample equilibrium fluctuations, may increase the importance of these descriptors in future studies.

\subsection{Comparison to brute-force BAR}

\begin{figure}[h!]
\centering
\begin{subfigure}{0.48\textwidth}
    \centering
    \caption{}
    \includegraphics[width=\linewidth]{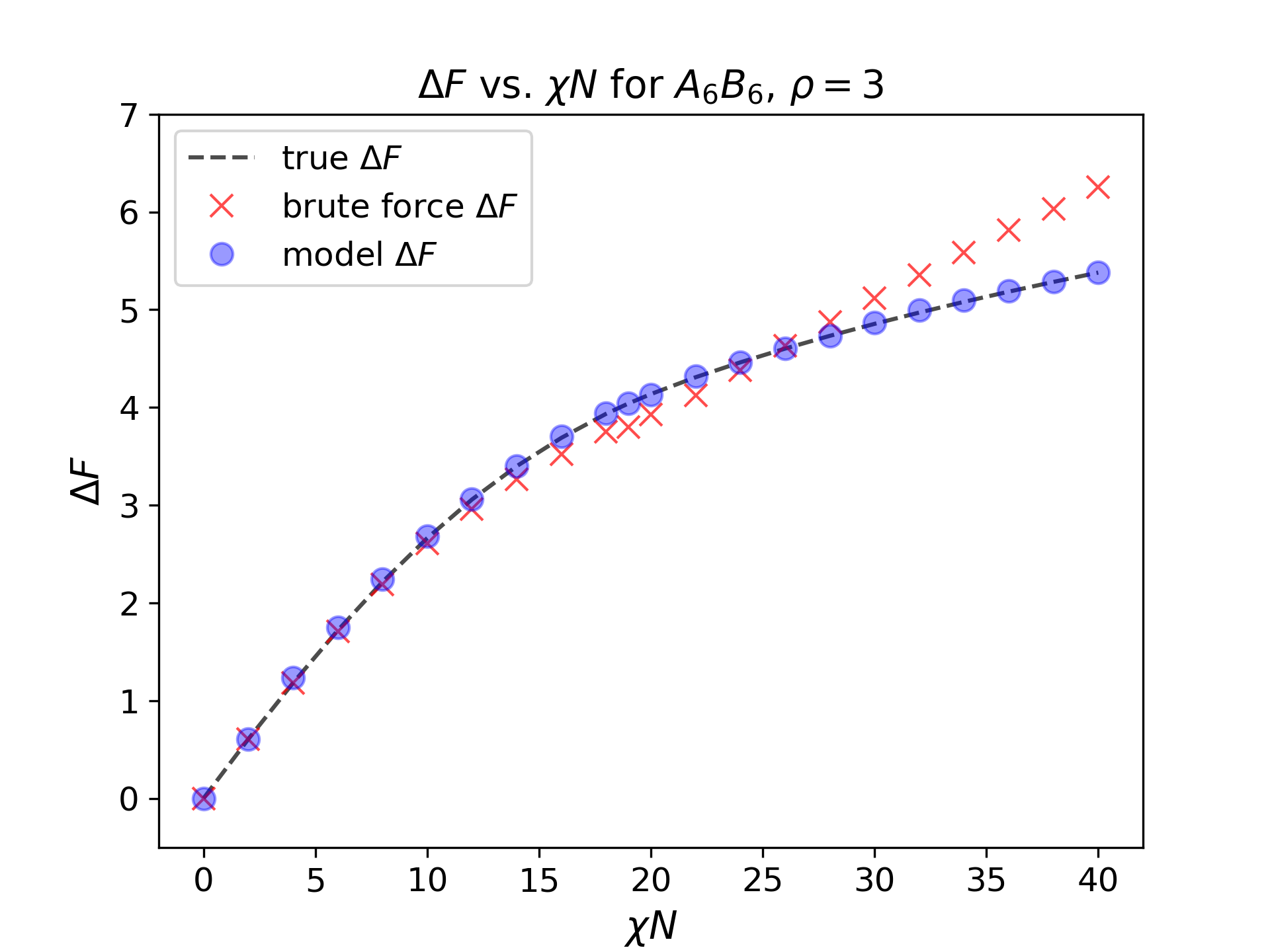}
\end{subfigure}
\begin{subfigure}{0.48\textwidth}
    \centering
    \caption{}
    \includegraphics[width=\linewidth]{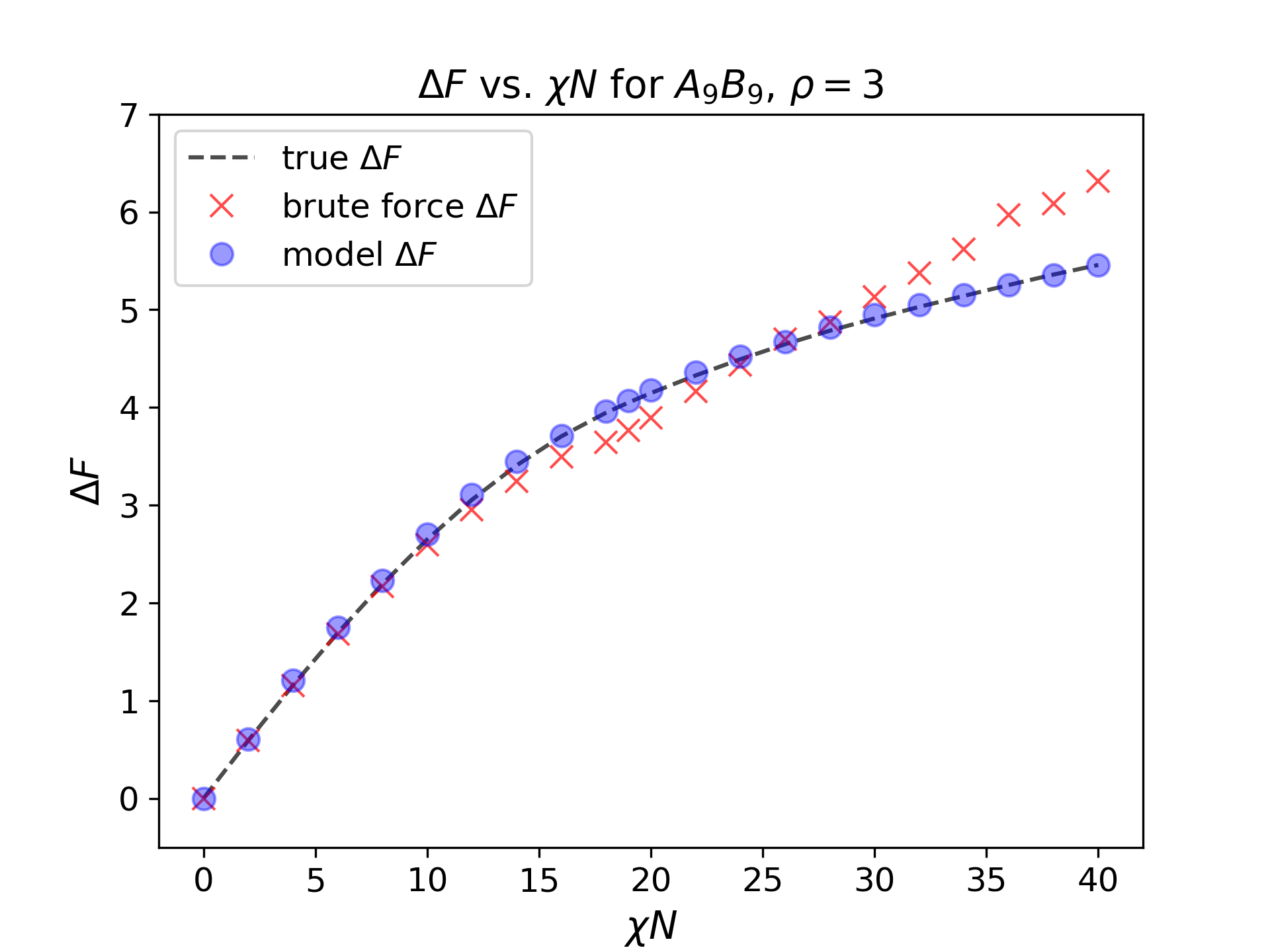}
\end{subfigure}

\vspace{5mm}

\begin{subfigure}{0.48\textwidth}
    \centering
    \caption{}
    \includegraphics[width=\linewidth]{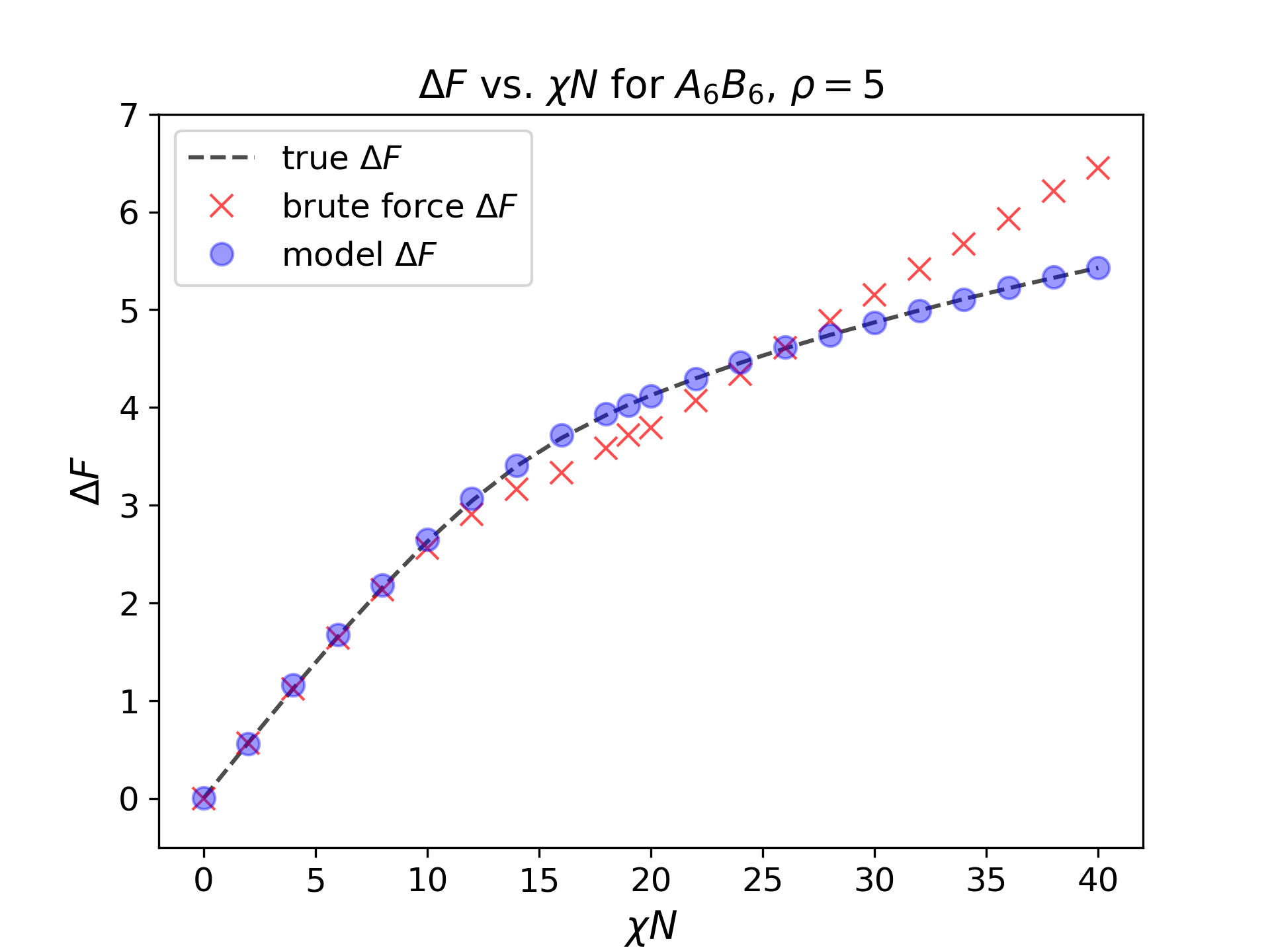}
\end{subfigure}
\begin{subfigure}{0.48\textwidth}
    \centering
    \caption{}
    \includegraphics[width=\linewidth]{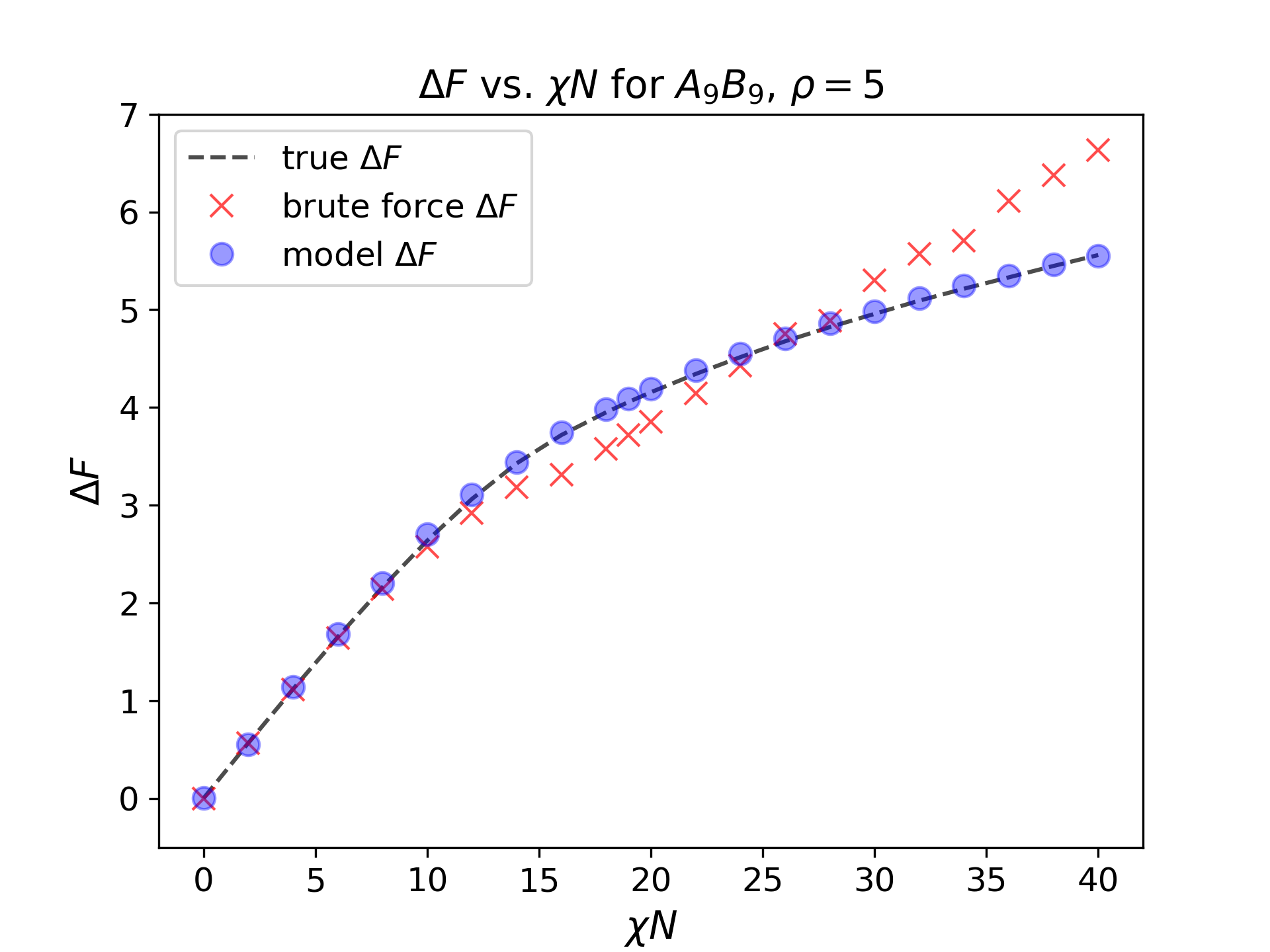}
\end{subfigure}
\caption{Free energy vs. $\chi N$ plots comparing average free-energy per chain values for the system for the optimized NN model vs. direct, brute-force BAR for 4 chains: (a) $A_6B_6$ and (b) $A_9B_9$ at $\rho = 3$; (c) $A_6B_6$ and (d) $A_9B_9$ at $\rho = 5$. }
\label{fig:bar_overlap}
\end{figure}

To further evaluate the predictive capability of the hyperparameter-optimized neural network model, we compare its free-energy predictions to estimates obtained using a direct, or ``brute-force,'' BAR calculation. In the standard procedure described previously, free energies are computed incrementally between neighboring states separated by $\Delta(\chi N)=2$. These intermediate free-energy differences are then summed sequentially from the reference state $\chi N=0$ to obtain $\Delta F$ at larger values of $\chi N$. Because each BAR calculation is performed between closely related thermodynamic states, the sampled configurational ensembles exhibit substantial phase-space overlap, ensuring accurate free-energy estimates.

In contrast, the brute-force BAR approach attempts to compute the free energy directly between the endpoints $\chi N=0$ and a target $\chi N$ without incorporating intermediate simulations. In this case the sampled ensembles correspond to thermodynamic states that may differ significantly in both interaction strength and polymer morphology. As $\chi N$ increases, these states share progressively less phase-space overlap, causing the BAR estimator to rely increasingly on rare configurations and leading to systematic inaccuracies in the resulting free-energy estimates. This comparison therefore provides a useful benchmark for assessing the extent to which the neural network model can reproduce accurate free energies in regimes where direct reweighting methods begin to fail.

The results of this comparison are shown in Fig.~\ref{fig:bar_overlap} for four polymer systems: $A_6B_6$ and $A_9B_9$ at bead densities $\rho=3$ and $\rho=5$. Each data point corresponds to an average across the 1000 polymer chains contained within the simulation box. Free energies were first computed on a per-chain basis using either the stratified BAR procedure (which we defined as the ground truth free energy) or the brute-force BAR estimator and subsequently averaged across chains to obtain the system-level values shown in the figure. The neural network predictions were obtained in an analogous manner: the trained model was evaluated independently for each chain and the predicted free energies were then averaged across the ensemble.\footnote{The standard errors of the chain-averaged free energies were extremely small ($<0.1$ for all systems considered), and the resulting error bars were therefore smaller than the marker size in the plots.}

Several trends are immediately apparent from Fig.~\ref{fig:bar_overlap}. First, the neural network predictions closely track the reference free energies obtained from the stratified BAR procedure across the entire range of $\chi N$ values considered. This agreement holds for both chain architectures examined in the testing dataset ($A_6B_6$, $N=12$ and $A_9B_9$, $N=18$) and for both simulated densities. The ability of the model to accurately reproduce the free-energy curves for these systems is particularly notable because neither chain architecture appears in the training dataset; instead, they belong to the held-out validation/testing set (Table~\ref{tab:chainset}). The results therefore indicate that the model successfully generalizes across chain lengths and densities rather than simply memorizing the training configurations.

In contrast, the brute-force BAR estimates increasingly deviate from the reference free energies as $\chi N$ grows. While the agreement is reasonable in the weak-segregation regime ($\chi N \lesssim 15$), the discrepancies become progressively larger at higher $\chi N$, where microphase separation and strong $A-B$ repulsion produce increasingly distinct configurational ensembles. This trend is consistent with the expected breakdown of phase-space overlap between the $\chi N=0$ reference state and highly segregated states. The deviations are particularly pronounced at $\rho=5$, where the brute-force estimates systematically overpredict the free energy at large $\chi N$. In comparison, the errors are somewhat smaller at $\rho=3$, suggesting that higher-density systems may exhibit stronger sensitivity to the loss of configurational overlap.

Taken together, these results demonstrate that the neural network model is capable of accurately reproducing the free energies obtained from the more reliable, but significantly more computationally expensive, stratified BAR procedure, even for chain architectures not included in the training set. At the same time, the comparison highlights the limitations of direct brute-force BAR calculations when applied to thermodynamic states separated by large changes in $\chi N$. In such regimes, the machine learning model provides a rapid and stable alternative for estimating free-energy differences without requiring intermediate simulations or repeated evaluation of the BAR equations.

\section{Concluding remarks}

In this work we presented a machine learning framework for rapidly predicting excess free energies of linear diblock copolymer (dBCP) systems using energetic descriptors obtained from simulation. Free energies are central quantities governing polymer phase behavior and thermodynamic stability, yet their computation typically requires extensive simulations and post-processing methods such as the Bennett Acceptance Ratio (BAR). While BAR provides highly accurate estimates when applied between closely related thermodynamic states, reliable evaluation across large changes in $\chi N$ requires a sequence of intermediate simulations to maintain sufficient phase-space overlap. This stratified procedure significantly increases the computational cost of constructing free-energy landscapes across polymer parameter spaces.

To address this challenge, we trained feed-forward neural networks to learn the relationship between per-bead energetic descriptors and excess free energies computed from dissipative particle dynamics simulations of freely-jointed chain (FJC) diblock copolymers. The input descriptors were derived directly from the physical interactions present in the model, including heterogeneous interaction energies, homogeneous interaction energies, and bonded spring energies. Time-averaged means, variances, and correlations of these quantities were used to represent both the typical energetic environment of each chain segment and the fluctuations associated with the accessible configurational ensemble. Model architectures were selected through hyperparameter optimization to identify networks capable of accurately learning the nonlinear mapping between these energetic descriptors and the corresponding free energies.

The resulting models reproduce the free energies obtained from the stratified BAR procedure with high accuracy across the range of chain architectures and densities considered. Importantly, the neural network predictions remain accurate for polymer systems belonging to the held-out testing dataset, demonstrating that the learned representation generalizes across chain lengths and simulation conditions rather than simply memorizing the training data. Comparisons with direct, brute-force BAR calculations further illustrate the advantage of the learned model: while brute-force BAR increasingly deviates from the reference free energies at large $\chi N$ due to insufficient phase-space overlap between ensembles, the neural network predictions remain consistent with the stratified BAR estimates. These results indicate that the model captures key thermodynamic relationships linking microscopic energetic distributions to macroscopic free-energy differences.

Rapid prediction of polymer free energies has several important implications. Such models can serve as efficient surrogates for computationally expensive free-energy calculations, enabling faster exploration of polymer parameter spaces and accelerating studies of block copolymer phase behavior. In the context of coarse-grained simulations, the framework developed here provides a practical tool for analyzing freely-jointed chain models and estimating thermodynamic quantities directly from sampled configurations without repeated evaluation of BAR equations or intermediate simulations.

Several promising directions for future work remain. Interpreting the trained models may provide new physical insight into the thermodynamics of polymer systems; techniques such as gradient-based sensitivity analysis or feature attribution methods (e.g., SHAP \cite{lundberg2017unified}) could reveal which energetic features most strongly influence predicted free energies and potentially suggest new theoretical relationships between microscopic energy fluctuations and macroscopic thermodynamic behavior. Additionally, extending the framework to nonequilibrium or partially equilibrated simulation data may allow machine learning models to rapidly assess whether polymer simulations have reached equilibrium. Finally, while the present study focuses on freely-jointed linear chains, the approach could be generalized to more complex polymer models. Extending the methodology to Gaussian chain representations and nonlinear polymer architectures, potentially through graph neural networks capable of encoding arbitrary molecular connectivity, would broaden the applicability of this framework to a wider class of polymer systems.

Overall, this work demonstrates that combining physically motivated simulation descriptors with modern machine learning models provides an effective strategy for accelerating the computation of polymer free energies while preserving the underlying thermodynamic interpretation of the problem.

\bibliographystyle{unsrt}
\bibliography{references}  

\end{document}